\documentclass[prd,preprintnumbers,showpacs,nofootinbib,12pt]{revtex4}
\usepackage{epsfig}
\usepackage{graphicx}
\usepackage {axodraw}

\usepackage{bm}
\usepackage{amsmath}
\usepackage{amssymb}
\usepackage{amscd}
\usepackage{latexsym}

\newcommand{\be}{\begin{equation}}

\newcommand{\ee}{\end{equation}}
\newcommand{\bea}{\begin{eqnarray}}
\newcommand{\eea}{\end{eqnarray}}

\newcommand{\nn}{\nonumber}

\def\Black{}
 \def\AliasBlue{}
 \def\Blue{}
 \def\Brown{}

\begin{document}

\newcommand{\bra}[1]{\langle #1|}
\newcommand{\ket}[1]{|#1\rangle}
\newcommand{\braket}[2]{\langle #1|#2\rangle}
\newcommand{\tr}{\textrm{Tr}}
\newcommand{\lag}{\mathcal{L}}
\newcommand{\mbf}[1]{\mathbf{#1}}
\newcommand{\desl}{\slashed{\partial}}
\newcommand{\Desl}{\slashed{D}}
%\preprint{{\bf DFF-413/05/04}}

\renewcommand{\bottomfraction}{0.7}
\newcommand{\epsi}{\varepsilon}

\newcommand{\nl}{\nonumber \\}
\newcommand{\tc}[1]{\textcolor{#1}}
\newcommand{\sla}{\not \!}
\newcommand{\spinor}[1]{\left< #1 \right>}
\newcommand{\cspinor}[1]{\left< #1 \right>^*}
\newcommand{\Log}[1]{\log \left( #1\right) }
\newcommand{\Logq}[1]{\log^2 \left( #1\right) }
\newcommand{\mr}[1]{\mathrm{#1}}
\newcommand{\cw}{c_\mathrm{w}}
\newcommand{\sw}{s_\mathrm{w}}
\newcommand{\ct}{c_\theta}
\newcommand{\st}{s_\theta}
\newcommand{\gt}{{\tilde g}}
\newcommand{\gtp}{{{\tilde g}^\prime}}
\renewcommand{\i}{\mathrm{i}}
\renewcommand{\Re}{\mathrm{Re}}
\newcommand{\yText}[3]{\rText(#1,#2)[][l]{#3}}
\newcommand{\xText}[3]{\put(#1,#2){#3}}

\def\to{\rightarrow}
\def\ptl{\partial}
\def\beq{\begin{equation}}
\def\eeq{\end{equation}}
\def\bea{\begin{eqnarray}}
\def\eea{\end{eqnarray}}
\def\nn{\nonumber}
\def\half{{1\over 2}}
\def\rhalf{{1\over \sqrt 2}}
\def\calo{{\cal O}}
\def\call{{\cal L}}
\def\calm{{\cal M}}
\def\del{\delta}
\def\eps{\epsilon}
\def\lam{\lambda}

\def\anti{\overline}
\def\delfac{\sqrt{{2(\del-1)\over 3(\del+2)}}}
\def\heff{h'}
\def\square{\boxxit{0.4pt}{\fillboxx{7pt}{7pt}}\hspace*{1pt}}
    \def\boxxit#1#2{\vbox{\hrule height #1 \hbox {\vrule width #1
             \vbox{#2}\vrule width #1 }\hrule height #1 } }
    \def\fillboxx#1#2{\hbox to #1{\vbox to #2{\vfil}\hfil}   }

\def\braket#1#2{\langle #1| #2\rangle}
\def\gev{~{\rm GeV}}
\def\gam{\gamma}
\def\sn{s_{\vec n}}
\def\sm{s_{\vec m}}
\def\mm{m_{\vec m}}
\def\mn{m_{\vec n}}
\def\mh{m_h}
\def\sumn{\sum_{\vec n>0}}
\def\summ{\sum_{\vec m>0}}
\def\vl{\vec l}
\def\vk{\vec k}
\def\ml{m_{\vl}}
\def\mk{m_{\vk}}
\def\gp{g'}
\def\gt{\tilde g}
\def\hw{{\hat W}}
\def\hz{{\hat Z}}
\def\ha{{\hat A}}

\def\yy{{\cal Y}_\mu}
\def\yyt{{\tilde{\cal Y}}_\mu}
\def\lq{\left [}
\def\rq{\right ]}
\def\dmu{\partial_\mu}
\def\dnu{\partial_\nu}
\def\dmus{\partial^\mu}
\def\dnus{\partial^\nu}
\def\gp{g'}
\def\gpt{{\tilde g'}}
\def\ggs{\frac{g}{\gs}}
\def\eps{{\epsilon}}
\def\tr{{\rm {tr}}}
\def\V{{\bf{V}}}
\def\W{{\bf{W}}}
\def\Wt{\tilde{ {W}}}
\def\Y{{\bf{Y}}}
\def\Yt{\tilde{ {Y}}}
\def\L{{\cal L}}
\def\s{s_\theta}
\def\st{s_{\tilde\theta}}
\def\c{c_\theta}
\def\ct{c_{\tilde\theta}}
\def\gt{\tilde g}
\def\et{\tilde e}
\def\At{\tilde A}
\def\Zt{\tilde Z}
\def\Wpt{{\tilde W}^+}
\def\Wmt{{\tilde W}^-}

\newcommand{\Apt}{{\tilde A}_1^+}
\newcommand{\Bpt}{{\tilde A}_2^+}
\newcommand{\Amt}{{\tilde A}_1^-}
\newcommand{\Bmt}{{\tilde A}_2^-}
\newcommand{\Wtp}{{\tilde W}^+}
\newcommand{\Atp}{{\tilde A}_1^+}
\newcommand{\Btp}{{\tilde A}_2^+}
\newcommand{\Atm}{{\tilde A}_1^-}
\newcommand{\Btm}{{\tilde A}_2^-}
\def\mathswitchr#1{\relax\ifmmode{\mathrm{#1}}\else$\mathrm{#1}$\fi}
\newcommand{\Pe}{\mathswitchr e}
\newcommand{\Pp}{\mathswitchr {p}}
\newcommand{\PZ}{\mathswitchr Z}
\newcommand{\PW}{\mathswitchr W}
\newcommand{\PD}{\mathswitchr D}
\newcommand{\PU}{\mathswitchr U}
\newcommand{\PQ}{\mathswitchr Q}
\newcommand{\Pd}{\mathswitchr d}
\newcommand{\Pu}{\mathswitchr u}
\newcommand{\Ps}{\mathswitchr s}
\newcommand{\Pc}{\mathswitchr c}
\newcommand{\Pt}{\mathswitchr t}
\newcommand{\rd}{{\mathrm{d}}}
\newcommand{\GW}{\Gamma_{\PW}}
\newcommand{\GZ}{\Gamma_{\PZ}}
\newcommand{\GeV}{\unskip\,\mathrm{GeV}}
\newcommand{\MeV}{\unskip\,\mathrm{MeV}}
\newcommand{\TeV}{\unskip\,\mathrm{TeV}}
\newcommand{\fba}{\unskip\,\mathrm{fb}}
\newcommand{\pba}{\unskip\,\mathrm{pb}}
\newcommand{\nba}{\unskip\,\mathrm{nb}}
\newcommand{\PT}{P_{\mathrm{T}}}
\newcommand{\PTmiss}{P_{\mathrm{T}}^{\mathrm{miss}}}
\newcommand{\CM}{\mathrm{CM}}
\newcommand{\inv}{\mathrm{inv}}
\newcommand{\sig}{\mathrm{sig}}
\newcommand{\tot}{\mathrm{tot}}
\newcommand{\backg}{\mathrm{backg}}
\newcommand{\evt}{\mathrm{evt}}
% particle masses
\def\mathswitch#1{\relax\ifmmode#1\else$#1$\fi}
\newcommand{\M}{\mathswitch {M}}
\newcommand{\R}{\mathswitch {R}}
\newcommand{\TEV}{\mathswitch {TEV}}
\newcommand{\LHC}{\mathswitch {LHC}}
\newcommand{\MW}{\mathswitch {M_\PW}}
\newcommand{\MZ}{\mathswitch {M_\PZ}}
\newcommand{\Mt}{\mathswitch {M_\Pt}}
\newcommand{\gs}{{g''}^2}
\def\lmu{{\bf L}_\mu}
\def\rmu{{\bf R}_\mu}
\def\si{\sigma}
\def\beqar{\begin{eqnarray}}
\def\eeqar{\end{eqnarray}}
\def\refeq#1{\mbox{(\ref{#1})}}
\def\reffi#1{\mbox{Fig.~\ref{#1}}}
\def\reffis#1{\mbox{Figs.~\ref{#1}}}
\def\refta#1{\mbox{Table~\ref{#1}}}
\def\reftas#1{\mbox{Tables~\ref{#1}}}
\def\refse#1{\mbox{Sect.~\ref{#1}}}
\def\refses#1{\mbox{Sects.~\ref{#1}}}
\def\refapps#1{\mbox{Apps.~\ref{#1}}}
\def\refapp#1{\mbox{App.~\ref{#1}}}
\def\citere#1{\mbox{Ref.~\cite{#1}}}
\def\citeres#1{\mbox{Refs.~\cite{#1}}}

\def\Black{}
 \def\AliasBlue{}
 \def\Blue{}
 \def\Brown{}

\vspace*{-20mm}
\begin{flushright}
SHEP-12-31
\end{flushright}
\vspace*{-2mm}

\title{Leptonic final states from di-boson production at the LHC\\ in the 4-Dimensional Composite Higgs Model}
 \author{D. Barducci$^*$, L. Fedeli$^*$ and S. Moretti}%
 \email{d.barducci,l.fedeli,s.moretti@soton.ac.uk}
 \affiliation{School of Physics and Astronomy, University of
 Southampton, Highfield,
 Southampton SO17 1BJ, UK}%
\author{S. De Curtis}%
 \email{decurtis@fi.infn.it}
 \affiliation{Istituto Nazionale di Fisica Nucleare, Sezione di Firenze,  Via G. Sansone 1, 50019 Sesto Fiorentino, Italy}%
 \author{G.M. Pruna}%
 \email{giovanni_marco.pruna@tu-dresden.de}
 \affiliation{TU Dresden, Institut ur Kern- und Teilchenphysik, \\
Zellescher Weg 19, D-01069 Dresden, Germany}%

\begin{abstract}
\noindent
We study di-boson production via both  neutral and charged current  at the Large Hadron Collider, i.e. subprocesses 
$q\bar q\to e^+\nu_e \mu^-\bar\nu_\mu$ + ${\rm{c.c.}}$
and 
$q\bar q'\to  l^+\nu_l~ l^{'+}l^{'-}$ +  ${\rm{c.c.}}$, respectively, 
where $q,q'$ are quarks and $l,l'=e,\mu$, in all possible combinations, in the context of the 4-Dimensional Composite Higgs Model.  These modes enable the production in the intermediate
steps of several additional -- with respect to the Standard Model -- neutral and charged gauge bosons belonging to the spectrum 
of this scenario, all of which in resonant topologies. We not only find these channels to be accessible over the background  but also show that, 
after a dedicated cut-based analysis, kinematic reconstruction of most such resonances is always possible. 
However, since the Electro-Weak precision data generally disfavor 
 neutral and charged gauge boson masses below the TeV range and also their
large  couplings
to light-fermions, these modes turn out to be relevant only for the 14 TeV option with high integrated luminosity. 
\end{abstract}

\pacs{12.60.Cn, 11.25.Mj, 12.39.Fe}% PACS, the Physics and Astronomy
\vspace*{-1.0mm}
\maketitle
\newpage
\section{Introduction}

In order to firmly establish the gauge sector of  the Standard Model (SM), i.e., masses and especially couplings of the
$\gamma, W^\pm$ and $Z$ gauge bosons, it would be sufficient to consider
Drell-Yan (DY) processes as well as di-boson hadro-production, both yielding leptonic final states (containing electrons and/or muons). 
On the one hand,
the DY processes enable one to access at once all the fermionic couplings of the SM, thanks to the fact that they are universal across 
generations.
On the other hand, di-boson hadro-production allows one to access the triple gauge self-couplings (for the latest experimental results, see the ATLAS \cite{Aad:2012rga} and CMS \cite{Chatrchyan:2012bd} papers), further recalling that the quartic ones are not gauge
independent {\it per se}.  Both statements remain true in any beyond the SM (BSM) scenario of electro-weak symmetry breaking (EWSB) 
where, no matter the gauge group describing the dynamics, such a universality assumption is maintained. In contrast, if the latter is 
dismissed, 
additional final states (involving heavy quarks and/or $\tau$ leptons) ought to be considered. 

Needless to say, the cleanliness of both DY and di-boson hadro-production, when searched for in $e,\mu$ final states, render them a 
favorite
from the experimental point of view: in general, the directions and energies of the particles of the emerging final states can be well reconstructed to the
extent that these scattering processes are ideal also for identifying the mass of the intermediate bosons being produced and 
studying their properties.
From the theoretical point of view, such mechanisms are well under control as higher order effects from both electro-weak (EW) interactions and 
Quantum Chromo-Dynamics (QCD), are well known and only affect the initial state (see, e.g., Ref.~\cite{Campbell:2006wx} for a review). 

In the presence of a Higgs-like signal, as testified by Large Hadron Collider (LHC) data recorded by the ATLAS~\cite{:2012gk} and 
CMS~\cite{:2012gu} collaborations, pathways towards BSM physics that incorporate a light scalar particle and encode possible deviations from the SM predictions should be considered with the highest priority.  A scalar particle
emerging from the Higgs mechanism  might not be the only means of generating masses for known (and possibly new) 
matter and force states, but an ingredient of a more general framework with new degrees of freedom and interactions appearing at the TeV scale. 
We adopt here the recently proposed 4-Dimensional Composite Higgs Model (4DCHM) of  Ref.~\cite{DeCurtis:2011yx}, 
which represents a complete and calculable scheme for the physics of the Higgs boson
as a pseudo-Nambu-Goldstone boson (PNGB), by capturing all the relevant features of 5D models and more in general of composite Higgs models based upon partial compositeness. 

Amongst the many BSM EWSB scenarios proposed over the years, the one with a  Higgs as a PNGB associated to the 
breaking of a strong underlying dynamics
yields one of the most natural solutions to the hierarchy problem of the 
SM \cite{Kaplan:1983fs,Georgi:1984ef,Georgi:1984af,Dugan:1984hq}. More recently such a scenario has been supplemented by the aforementioned mechanism of 
partial compositeness. The simplest example, based on $SO(5)/SO(4)$ is considered in \cite{Agashe:2004rs}, of which the scheme proposed in \cite{DeCurtis:2011yx} represents a highly deconstructed 4D version. The latter represents the
 minimal choice for the ensuing enlarged (and composite) fermionic sector,  leading to the minimal matter content that allows for a finite Higgs potential calculable via the Coleman-Weinberg technique
and includes a set of degrees of freedom (both bosonic and fermionic) which might well be accessible at the LHC. From the 
explicit expression of the Higgs potential  in \cite{DeCurtis:2011yx}, one can then extract  the Higgs vacuum expectation value (VEV) and mass in terms of the model parameters.  
The peculiarity of this BSM scenario is that, for a natural choice of the free parameters both in  the gauge and fermion sectors, 
the spectrum of the composite Higgs masses that one obtains includes values that are compatible with the most recent LHC results. In particular,  the request of reproducing a Higgs mass in the vicinity of 125 GeV, due to the correlation between the Higgs mass value and the one of the lightest new fermionic states \cite{Matsedonskyi:2012ym,Pomarol:2012qf,Redi:2012ha,Marzocca:2012zn,Panico:2012uw}, implies the presence of new fermions in the TeV range (or below) so that they might be within the reach of the present and/or future runs of LHC.

The study of the composite Higgs and fermionic sectors of such a class of models has received much attention over the last few years \cite{RediTalk}.
Here, we are instead concerned with the gauge sector. In fact, the latter is also extremely rich in general 
as it predicts  extra composite spin-1 resonances. In particular, in the formulation
of Ref.~\cite{DeCurtis:2011yx} (see the next Section for details), there are five extra $Z'$ states and three extra $W'$ states. These 
objects are weakly yet sizably coupled to the first and second generations of fermionic matter (in both the quark and lepton sector) and this makes the 
4DCHM an excellent candidate for a phenomenological analysis of DY and di-boson processes at the LHC.  However, we remark that such a theoretical set up realizes partial compositeness only for the third generation of quarks\footnote{Following a minimal approach, the partial compositeness for the third generation leptons is not embedded in the model.} of the SM, hence in principle one should also rely on the study of $b\bar b$ and $t\bar t$ final states (in addition to the leptonic ones) in order to extract from data the complete structure of the new EW sector.

While the study of the hadronic final states in both DY and di-boson channels will be pursued in a separate publication \cite{Ken:2012pp}
and the one of the leptonic ones in DY processes has been tackled in Ref.~\cite{Barducci:2012kk},  it is the purpose of this paper to
 investigate, in the context of the aforementioned 4DCHM,
the phenomenology of both charged
\begin{equation}\label{eq:processWW}
pp(q\bar q)\to W^+W^-\to e^+\nu_e \mu^-\bar\nu_\mu~+~{\rm{c.c.}}\to e^\pm\mu^\mp E^{T}_{\rm miss}
\end{equation}
and mixed
\begin{equation}\label{eq:processWZ}
pp(q\bar q')\to W^\pm Z\to  l^+\nu_l~ l^{'+}l^{'-}~+~{\rm{c.c.}}\to l^\pm  l^{'+}l^{'-}  E^{T}_{\rm miss}
\end{equation}
di-boson production at the LHC, yielding different-flavor opposite-charge di-leptons plus missing transverse energy in the first 
case (henceforth, the $2l$ signature) 
and all-flavor ($l,l'=e,\mu$) and charge tri-leptons  plus missing transverse energy in the second
case (henceforth, the $3l$ signature)\footnote{ Note that di-lepton final states with identical flavors are of no use for the  (\ref{eq:processWW}) process, 
as they are burdened by an overwhelming 
SM background induced by $ZZ$ events, with one $Z$ boson decaying invisibly. Note also that the contribution of process (\ref{eq:processWZ}) to the 
$2l$ signature, occurring when a lepton escapes the detector, was found to be negligible in Ref.~\cite{DeCurtis:2012cn}.}. 
 In both instances, which at times
we refer to as the $WW$ and $WZ$ channels, 
the symbols $W^\pm$ and $Z$ refer to any possible charged and neutral, respectively, spin-1 massive gauge bosons present in the 4DCHM 
whilst
$q(\bar q)$ and $q'(\bar q')$ are the (anti)quarks found inside the proton. 

The plan of the paper is as follows. In the next Section, \ref{sec:4DCHM}, we briefly sketch the salient features of the gauge sector of 
the 4DCHM.
In Sect.~\ref{sec:results}, after a short description of the computational tools used, we present our results for the two channels  
(\ref{eq:processWW})--(\ref{eq:processWZ}). We conclude in 
Sect.~\ref{sec:summa}.  Appendix A contains instead some numerical values of 4DCHM couplings entering our phenomenological analysis.

%%%%%%%%%%%%%%%%%%%%%%%%%%%%%%%%%%%%%%%%%%%%%%%%%%%
%%%%%% SECTION MODEL DESCRIPTION %%%%%%%%%%%%%%%%%%
%%%%%%%%%%%%%%%%%%%%%%%%%%%%%%%%%%%%%%%%%%%%%%%%%%%

\section{The gauge sector of the 4DCHM}
\label{sec:4DCHM}

In this Section we describe the main characteristics of the gauge sector of the 4DCHM introduced in \cite{DeCurtis:2011yx}, 
where further details can be found,  which is based on a low-energy Lagrangian approximation of the deconstructed 5D Minimal Composite
Higgs Model (MCHM) introduced in \cite{Agashe:2004rs}, based on the coset $SO(5)/SO(4)$ with four Goldstone bosons, the latter containing the Higgs state.

The 4DCHM can be schematized in two sectors, the elementary and the composite one, arising from an extreme deconstruction of the 5D theory.
This two-site truncation represents the framework where to study both bosonic and fermionic new resonances that might be accessible at the LHC, 	
though it captures all the relevant features of the original MCHM with the Higgs boson arising as a PNGB.
The gauge structure of the elementary sector of the 4DCHM is associated with the $SU(2)_L\otimes U(1)_Y$ SM  gauge symmetry whereas the composite sector has a local $SO(5)\otimes U(1)_X$  symmetry
that gives rise to eleven new gauge resonances.
Therefore, the spin-1 particle content of the 4DCHM is given, besides the standard $W,~Z$ bosons and the photon, by five new neutral, collectively denoted by $Z'$, and three new charged, collectively denoted by $W'$, bosons. The parameters for the gauge sector are obtained from the scale $f$ of the spontaneous global symmetry breaking $SO(5)\to SO(4)$  (typically of the order of 1 TeV) and $g_*$, the $SO(5)$ gauge coupling constant which, for simplicity, we take equal to the $U(1)_X$ one. The mass spectrum of the spin-1 fields is then expressed in terms of these two new parameters and the SM ones: $g_0$ and $g_{0Y}$, the gauge couplings of $SU(2)_L$ and $U(1)_Y$ respectively.  The analytical expressions of the gauge boson masses at the leading order in $\xi=v^2/f^2$,  with $v$ the VEV of the Higgs, are given in eqs.
 (\ref{eq:MW})--(\ref{eq:MZ}), where an increasing
number in the label  indicates a particle with higher mass.

For the charged sector we have:
\begin{eqnarray}
\label{eq:MW}
M^2_{W}\simeq && \frac{f ^2}{4} g_*^2s^2_\theta  \xi,\nonumber\\
M^2_{W_1}=&& f ^2g_*^2,\nonumber\\
M^2_{ W_2}\simeq&& \frac{f ^2g_*^2}{ c_\theta^2}(1-\frac{s^2_\theta c^4_\theta}{2 c_{2\theta}}\xi),\nonumber\\
M^2_{W_3}\simeq&&2 f ^2g_*^2 (1- \frac{s^2_\theta}{4c_{2\theta}}\xi),
\end{eqnarray}
with $\tan\theta=(s_\theta/c_\theta)=(g_0/g_*)$. Note that $W_1$ mass does not receive any contribution from  EWSB.
 
For the neutral sector we  get:
\begin{eqnarray}\label{eq:MZ}
M^2_{\gamma}= &&0,\nonumber\\
M^2_{Z}\simeq && \frac{f ^2}{4} g_*^2(s^2_\theta+\frac{s^2_\psi}{2})  \xi,\nonumber\\
M^2_{Z_1}=&& f ^2g_*^2,\nonumber\\
M^2_{ Z_2}\simeq&& \frac{f ^2g_*^2}{ c_\psi^2}(1-\frac{s^2_\psi c^4_\psi }{4 c_{2\psi}}\xi),\nonumber\\
M^2_{Z_3}\simeq&& \frac{f ^2g_*^2}{ c_\theta^2}(1-\frac{s^2_\theta c^4_\theta }{4 c_{2\theta}}\xi), \nonumber\\
M^2_{Z_4}=&&2 f ^2g_*^2, \nonumber\\
M^2_{Z_5}\simeq&& 2 f ^2g_*^2(1+\frac 1 {16} (\frac 1 {c_{2\theta}}+\frac 1{2 c_{2 \psi}})\xi)
\end{eqnarray}
with $\tan\psi=(s_\psi/c_\psi)=(\sqrt{2} g_{0Y}/g_*)$. The photon is massless, as it should be, and the neutral gauge bosons $Z_1$  and $Z_4$ have their masses completely determined by the composite sector. 

As stated before, in the 4DCHM  the VEV of the Higgs $v$ is extracted by the minimum of the Coleman-Weinberg potential as a function of the fermion and gauge boson parameters, which, in the following analysis, will be chosen in such a way as to reproduce $v=246$ GeV
 (see \cite{Barducci:2012kk} for details).  Eqs. (\ref{eq:MW})--(\ref{eq:MZ}) explicitly show the leading corrections to the mass spectrum due to EWSB. 

Regarding the fermionic sector, we just recall that the new heavy fermions are embedded in  fundamental representations of $SO(5)\otimes U(1)_X$ and two multiplets of resonances for each of the SM third generation quark are introduced in such a way that only top and bottom quarks
 mix with these heavy fermionic resonances in the spirit of partial compositeness. For processes (\ref{eq:processWW}) and (\ref{eq:processWZ})  we only need the 
the couplings of the $Z'$ and $W'$ (to which we refer in the following also explicitly as $Z_{i=1,...,5}$ and $W_{i=1,2,3}$) to the first two generations of leptons and quarks which live in the elementary sector. These couplings come from the mixing of the  $Z'$  and $W'$ with the elementary gauge bosons
which, in turn, see their couplings modified due to the same mixing.
In order to give an idea of the order of magnitude of this effect, we provide here the analytical expression for the charged and neutral current interaction Lagrangian for the 4DCHM at the leading order in $\xi$ (all the forthcoming calculations of cross sections are performed numerically though, with the corresponding full expressions without any approximations).  

For the charged-current Lagrangian we have:
\begin{equation}
{\cal L}_{CC}=[ g_W^+  W^+ +g_{W_1}^+ W_1^++ g_{W_2}^+ W_2^++ g_{W_3}^+ W_3^+ ]J^- +h.c.
\label{LCC}
\end{equation}
with $J^\pm=(J^1\pm i J^2)/2$, $ J^i_\mu=\bar\psi T^i_L \gamma_\mu[(1-\gamma_5)/2]\psi$,  and

\begin{eqnarray}
g_W^\pm=&&-\frac{g_* s_\theta}{\sqrt 2}(1+\frac  {s_\theta}{4 c_\theta}  a_{12}  \xi)\label{Wff},\\
g_{W_1}^\pm= &&0\label{W1ff},\\
g_{W_2}^\pm= &&\frac{g_* s_\theta^2}{\sqrt{2}c_\theta}    (1+\frac 1 4 (a_{22}-\frac{ c_\theta}{s_\theta}a_{12})\xi )\label{W2ff},\\
 g_{W_3}^\pm= &&\frac{g_* s_\theta^2}{2\sqrt{2}c_\theta} a_{24}\sqrt{\xi},
\label{W3ff}
\end{eqnarray}
where 
\begin{equation}
a_{12}= -\frac 1 4 c_\theta (1-4 c_\theta^2) s_\theta,  \quad \quad
a_{22}=-\frac{ c_\theta^2}{4(1-2 c_\theta^2)^2}, \quad \quad
a_{24}=-\frac{c_\theta }{\sqrt{2}(1-2c_\theta^2)}.
\end{equation}
As it is clear from eqs.~(\ref{W1ff}) and  (\ref{W3ff})$, W_1^\pm$ is fermiophobic and  $W_3^\pm$ is weakly coupled (this conclusion does not refer to the third generation quarks where the mixing with the composite fermions must be taken into account).  In Appendix A we list the numerical values for these couplings for the benchmark points 
used in our numerical analysis.

Analogously we derive the neutral-current  Lagrangian. Starting from the elementary sector, where the neutral gauge fields of $SU(2)_L\times U(1)_Y$ are coupled with the fermion currents, we get, after taking into account the mixing among the fields, the following expression:
\begin{equation}\label{LNC}
{\cal L}_{NC}=\sum_f\big[ e\bar\psi^f \gamma_\mu Q^f \psi^f A^\mu+ \sum_{i=0}^5   (\bar\psi^f_L  g_{Z_i}^L(f) \gamma_\mu  \psi^f_L+\bar\psi^f_R  g_{Z_i}^R(f) \gamma_\mu  \psi^f_R ) Z_i^\mu \big], 
\end{equation}
where $\psi_{L,R}=[(1\pm\gamma_5)/2]\psi$ and we have identified $Z_0$ with the neutral SM gauge boson $Z$. The photon field, $A_\mu$, is coupled to the electromagnetic current in the standard way, i.e., with
\begin{equation}\label{e}
e=\frac{g_L g_Y}{\sqrt{g_L^2+g_Y^2}},\quad\quad g_L=g_0 c_\theta,\quad \quad g_Y=g_{0Y} c_\psi ,
\end{equation}
while the couplings of the $Z_i$'s have the following expressions:
\begin{equation}
g_{Zi}^L(f)= A_{Z_i}T^3_L(f)+ B_{Z_i} Q^f, \quad\quad
g_{Zi}^R(f)=  B_{Z_i}Q^f,
\end{equation}
where $A_{Z_i}=(g_0 \alpha_i - g_{0Y} \beta_i) $, $B_{Z_i}=g_{0Y} \beta_i$,  
with $\alpha_i$ and $\beta_i$ the diagonalization matrix elements, namely: 
\begin{equation}
W_3=\sum_{i=0}^5 \alpha_i Z_i, \quad\quad  Y=\sum_{i=0}^5 \beta_i Z_i.
\end{equation}
Here $W_3$ and  $Y$ are the elementary gauge field associated to $SU(2)_L$ and $U(1)_Y$, respectively.
As a result, the $Z_1$ and $Z_4$ bosons are not coupled to leptons and to the first two quark generations, so they are completely inert for the processes we are here considering.

At the leading order in $\xi$ we get:
%cambiato segno ai coupling Z2
\begin{eqnarray}
&&A_{Z}= \sqrt{g_L^2+g_Y^2}\big[1+(\frac{g_L^2}{g_L^2+g_Y^2} a_Z+ \frac{g_Y^2}{g_L^2+g_Y^2} b_Z)\xi\big],~
B_{Z}= -\frac{g_Y^2}{\sqrt{g_L^2+g_Y^2}}(1+b_Z\xi), \\
&&A_{Z_2}=  -g_Y \frac{s_\psi}{c_\psi} \Big[1+(\frac{g_L}{g_Y} a_{Z_2}-b_{Z_2})\xi\Big], \quad\quad  B_{Z_2}= g_Y \frac{s_\psi}{c_\psi} \Big[1-b_{Z_2}\xi\Big],\\
&&A_{Z_3}=  -g_L\frac{s_\theta}{c_\theta}\big[1+(a_{Z_3}+\frac{g_Y}{g_L} b_{Z_3})\xi\big], \quad \quad   B_{Z_3}=   g_Y \frac{s_\theta}{c_\theta} b_{Z_3}\xi,\\
&&A_{Z_5}= (g_L a_{Z_5}-g_Y b_{Z_5})\sqrt{\xi},\quad\quad   B_{Z_5}=  g_Y b_{Z_5}\sqrt{\xi},
\end{eqnarray}
with
\begin{eqnarray}
a_{Z}= && (2 s_\theta^2+s_\psi^2)(4 cˆ_\theta^2-1)/32,\quad \quad
b_{Z}=  (2 s_\theta^2+s_\psi^2)(4 cˆ_\psi^2-1)/32, \\
a_{Z_2}=&& \frac{\sqrt{2} s_\theta s_\psi c_\psi^6}{4(c_\psi^2-c_\theta^2)(2 c_\psi^2-1)} ,     \quad \quad
b_{Z_2}= \frac{c_\psi^4(2-7c_\psi^2+9 c_\psi^4-4 c_\psi^6)}{8s_\psi^2 (1-2 c_\psi^2)^2},\\
a_{Z_3}= &&  \frac{-2 c_\theta^4+5 c_\theta^6-4 c_\theta^8}{4(1-2 c_\theta^2)^2},\quad \quad
b_{Z_3}=  \frac{\sqrt{2} s_\theta s_\psi c_\theta^6}{4 (2 c_\theta^2-1)(c_\theta^2-c_\psi^2) },\\
a_{Z_5}= &&  \frac{s_\theta}{2 \sqrt{2}(1-2 c_\theta^2)},\quad \quad
b_{Z_5}= - \frac{s_\psi}{4(1-2 c_\psi^2)}.
\end{eqnarray}
The numerical values of these fermionic couplings for  two benchmark points of our analysis are given in 
Appendix A.

The mixing among  the gauge bosons of the 4DCHM also leads to tri-linear and quadri-linear interactions between the heavy and  light gauge bosons. Their analytical expressions are quite complicated (also at the leading order in $\xi$), so we will not report them here, but again, we give in Appendix A the numerical values of the tri-linear gauge couplings (only) for
the case of the benchmarks used in the upcoming production and decay analysis.

Despite the large number of parameters in the fermionic sector, limited
to the analysis that we are going to perform in this paper, we can easily summarize its main   characteristics.
As already known (see, e.g., \cite{Matsedonskyi:2012ym}, \cite{Pomarol:2012qf} and \cite{Redi:2012ha} for a review) in models where the Higgs is a light composite PNGB the latter
is associated with light fermionic resonances (with both standard and exotic quantum numbers) around the mass of 1 TeV. We now stress that, as pointed out in
\cite{Barducci:2012kk}, for the purpose of this paper is sufficient to divide the composite fermion mass spectrum  in two different regimes, as follows.
\begin{itemize}
\item A regime where the mass of the lightest fermionic resonance is too heavy to allow for the decay of a $Z'$ and/or a $W'$ in a pair of heavy fermions and,
consequently, the widths of the $Z'$ and/or $W'$ are small, typically well below $100$ GeV. This configuration of the 4DCHM is illustrated by the forthcoming
benchmarks (a)--(f).
\item A regime where a certain number of masses of the new fermionic resonances are light enough to allow for the decay 
of a $Z'$ and/or $W'$ in a pair of heavy fermions and, consequently, the widths of the involved $Z'$ and/or $W'$ states are relatively large and can become
even comparable with the masses themselves. This configuration of the 4DCHM is illustrated by the forthcoming {\it colored} benchmarks
 (red, green, cyan, magenta, black, yellow).
\end{itemize}
 
In scanning the 4DCHM parameter space, we have of course checked that the regions eventually investigated via processes 
(\ref{eq:processWW})--(\ref{eq:processWZ}) are compatible with  LHC direct searches for heavy gauge bosons \cite{Aad:2011fe,Chatrchyan:2012qk,Hayden:2012gc,Chatrchyan:2012it} and fermions \cite{Aad:2012bb,ATLAS:2012aw,Chatrchyan:2012fp}.
Further, the top, bottom and Higgs masses emerging in the 4DCHM are limited as follows:
 $165 ~{\rm GeV} \le m_t \le 175 ~{\rm GeV}$,  $2 ~{\rm GeV} \le m_b \le 6 ~{\rm GeV}$ and  $124 ~{\rm GeV} \le m_H \le 126 ~{\rm GeV}$, the
latter consistent with the recent data coming from the ATLAS \cite{:2012gk} and CMS \cite{:2012gu} experiments.  We use $e,M_Z,G_F$ as input to further constraint the 4DCHM parameter space. However, regarding EWPTs, it should be mentioned here 
that,
as it is well known, extra gauge bosons give a positive contribution to the Peskin-Takeuchi $S$ parameter and the requirement of consistency with the EWPTs generally gives a bound on the mass of these resonances around few TeV \cite{Marzocca:2012zn}. 
In contrast, the fermionic sector is quite irrelevant for $S$ since the extra fermions are weakly coupled to the SM gauge bosons. Either way, as noticed in \cite{Contino:2006qr}, when dealing with effective theories, one can only parametrize $S$ rather than calculating it. In other words, since we are dealing with a truncated theory describing only the lowest-lying resonances that may exist, we need to invoke an Ultra-Violet (UV) completion for the physics effects we are not including in our description. These effects could well compensate for $S$, albeit with some tuning. One example is given in \cite{Contino:2006qr} by considering the contribution of higher-order operators in the chiral expansion. Another scenario leading to a reduced $S$ parameter is illustrated in \cite{DeCurtis:2011yx}, by including non-minimal interactions in the 4DCHM. For these reasons, in the following phenomenological analysis we will choose values for the gauge boson resonance masses around 2 TeV to avoid big contributions to the $S$ parameter. This choice corresponds to a compositeness scale $f$ around 1 TeV and  $g_*$ around 2. 

\begin{figure}[!ht]
\begin{center}
\vspace{-.8cm}
\unitlength1.0cm
\begin{picture}(7,10)
\put(-4.3,2.7){\epsfig{file=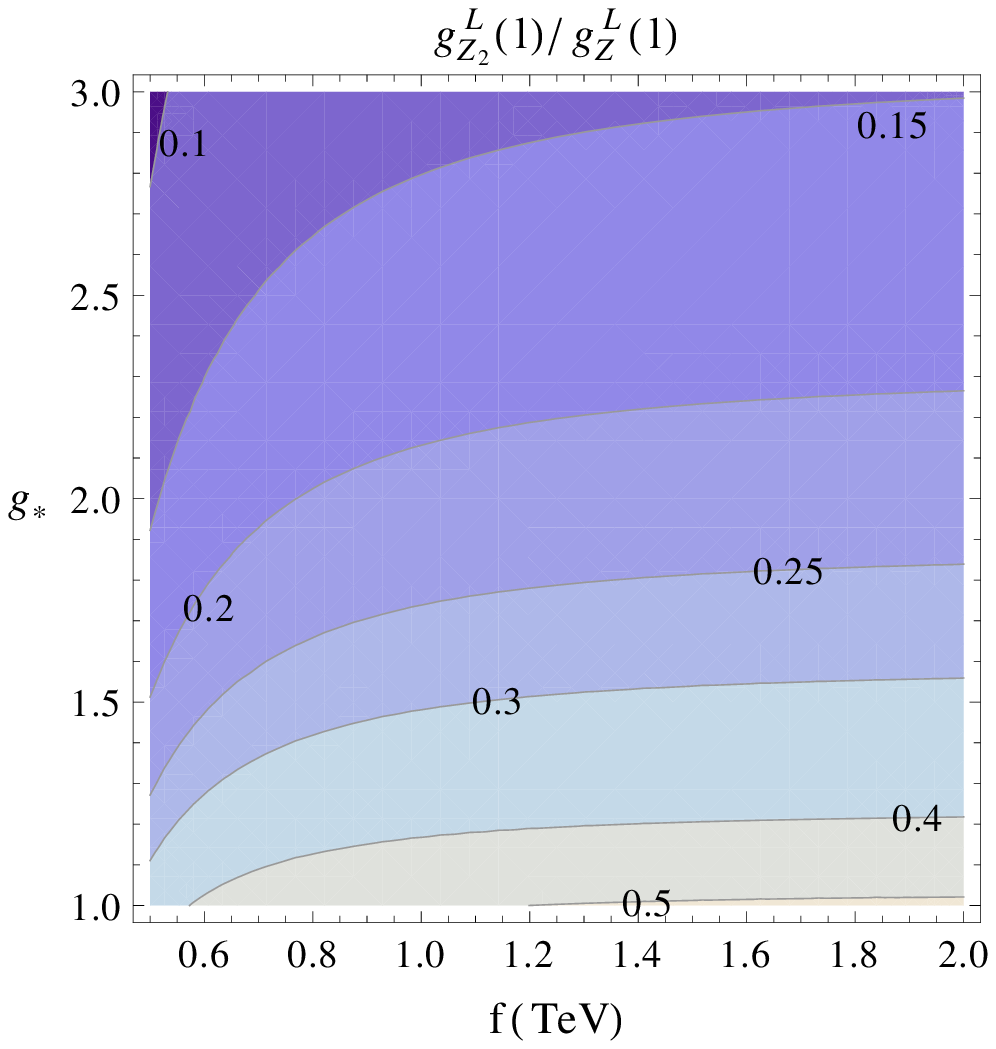,width=7.5cm}}
\put(3.5,2.7){\epsfig{file=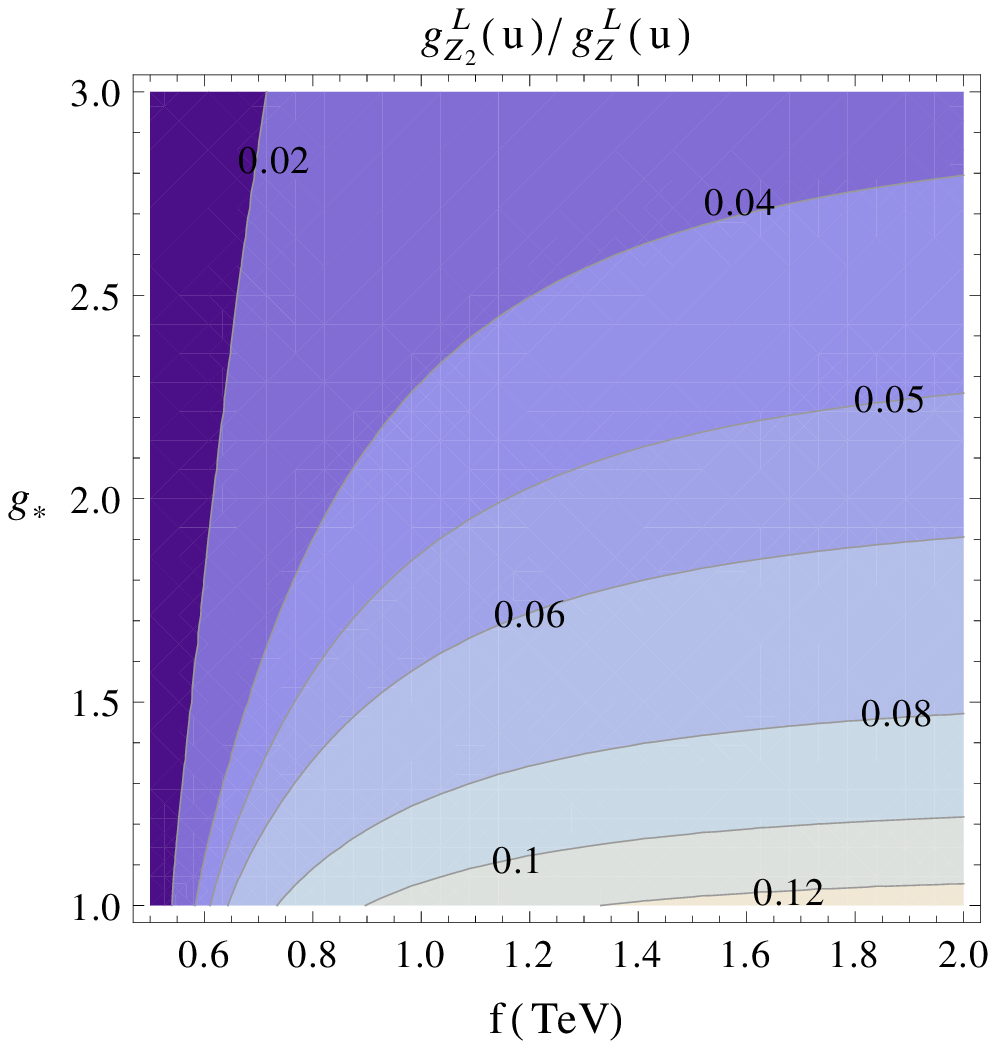,width=7.5cm}}
\put(-4.3,-5.3){\epsfig{file=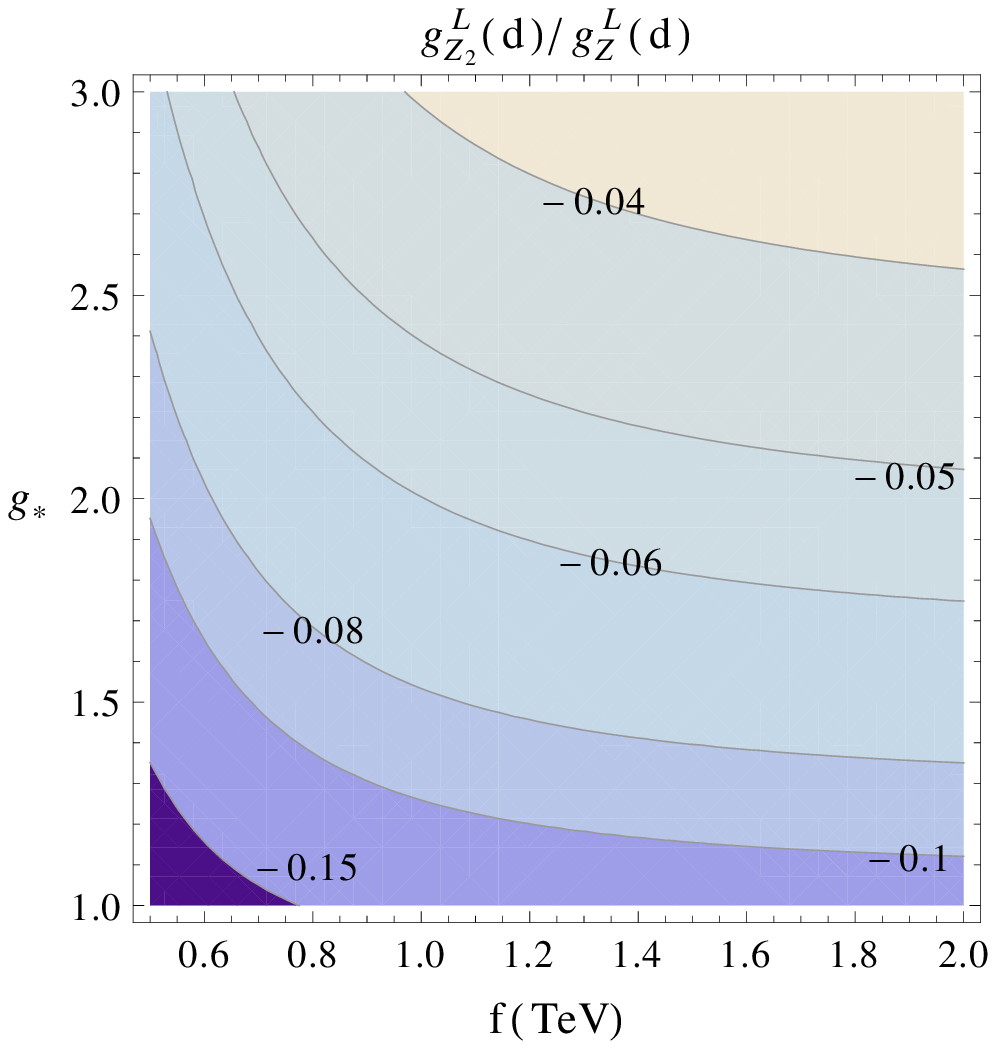,width=7.5cm}}
\put(3.5,-5.3){\epsfig{file=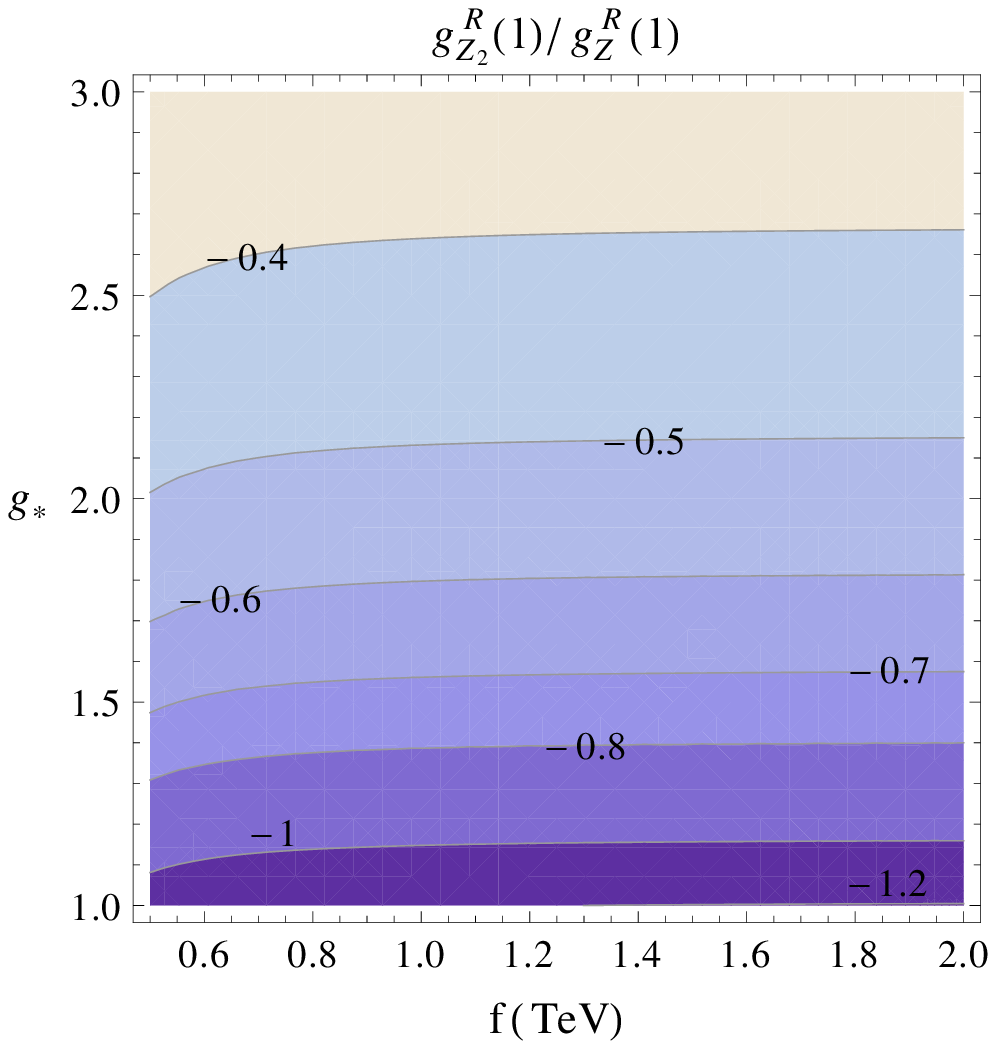,width=7.5cm}}
\end{picture}
\end{center}
\vskip 5.cm \caption{Ratio between the $Z_2$  light-fermion couplings and the $Z$ ones.
Top-left: for $g_{Z_2}^L(l)/g_{Z}^L(l)$. Top-right: for $g_{Z_2}^L(u)/g_{Z}^L(u)$.
Bottom-left: for $g_{Z_2}^L(d)/g_{Z}^L(d)$. Bottom-right: for $g_{Z_2}^R(l)/g_{Z}^R(l)$.}
\label{fig:Z2-femions}
\end{figure}

\begin{figure}[!ht]
\begin{center}
\vspace{-.8cm}
\unitlength1.0cm
\begin{picture}(7,10)
\put(-4.3,2.7){\epsfig{file=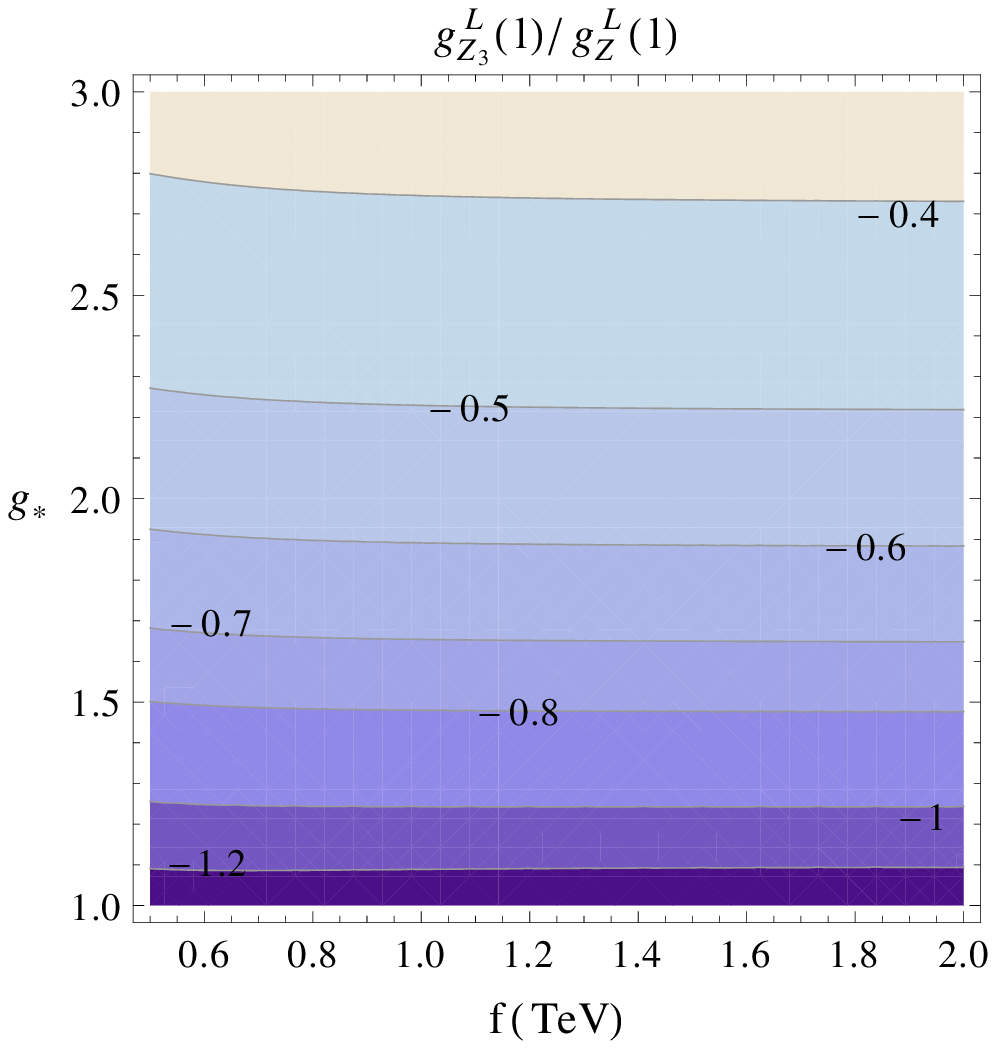,width=7.5cm}}
\put(3.5,2.7){\epsfig{file=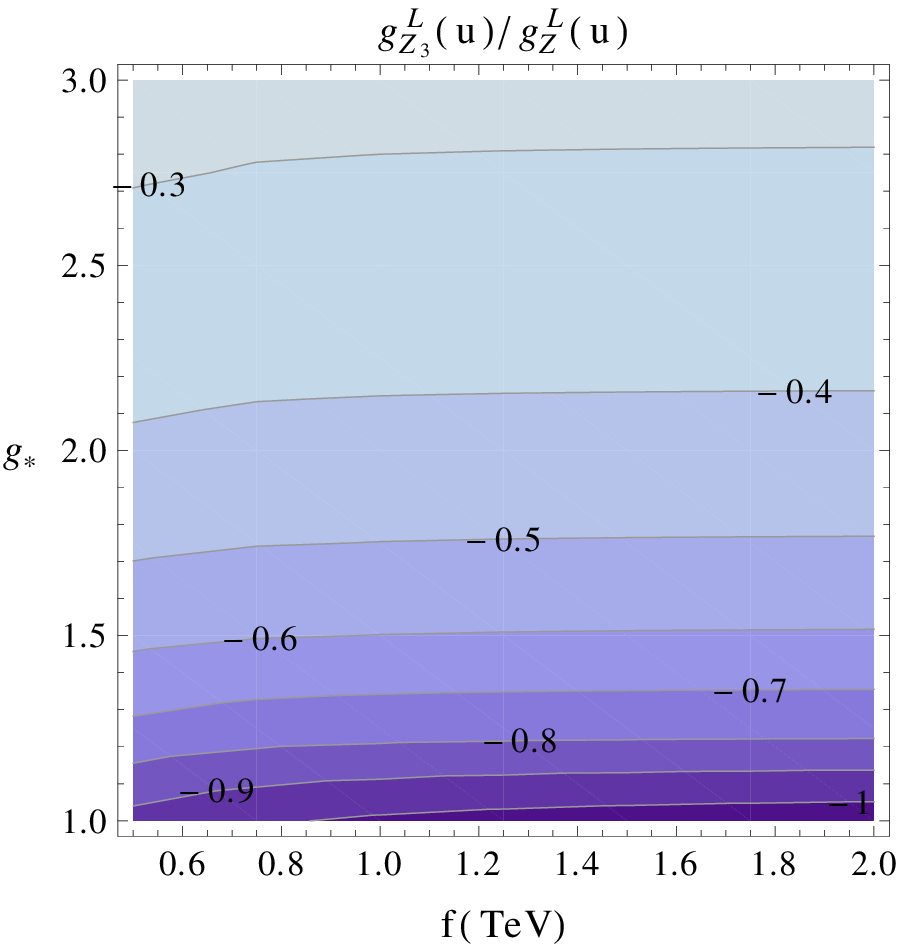,width=7.5cm}}
\put(-4.3,-5.3){\epsfig{file=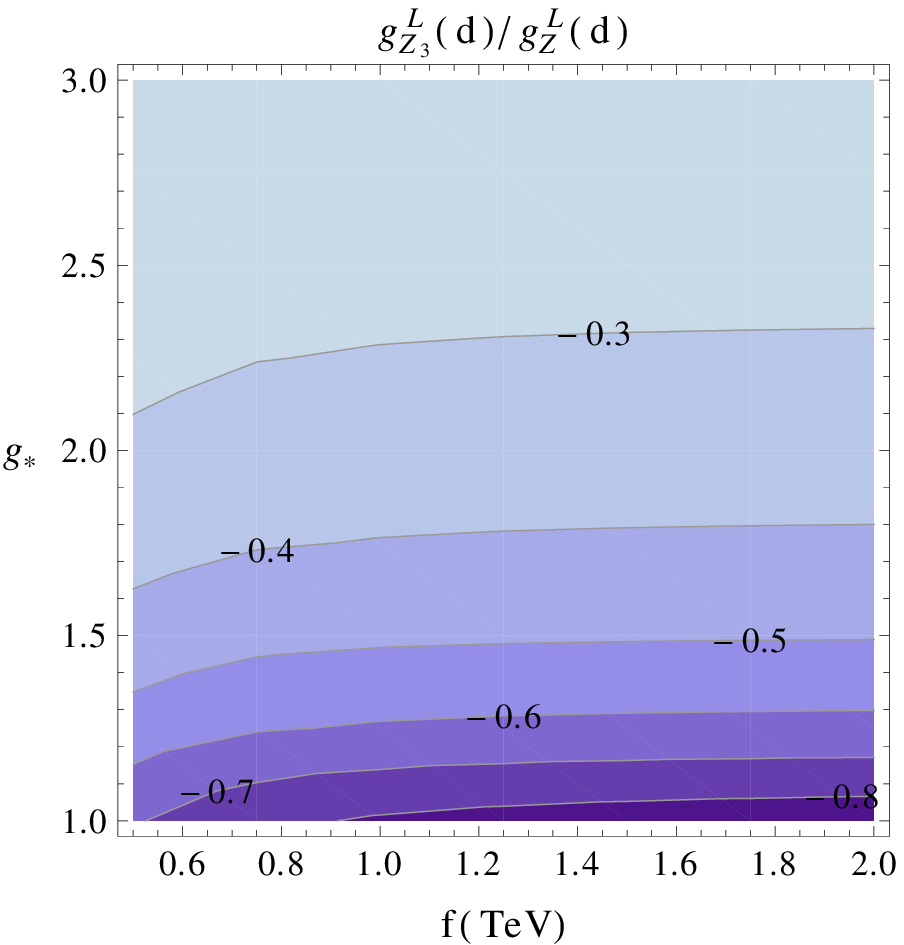,width=7.5cm}}
\put(3.5,-5.3){\epsfig{file=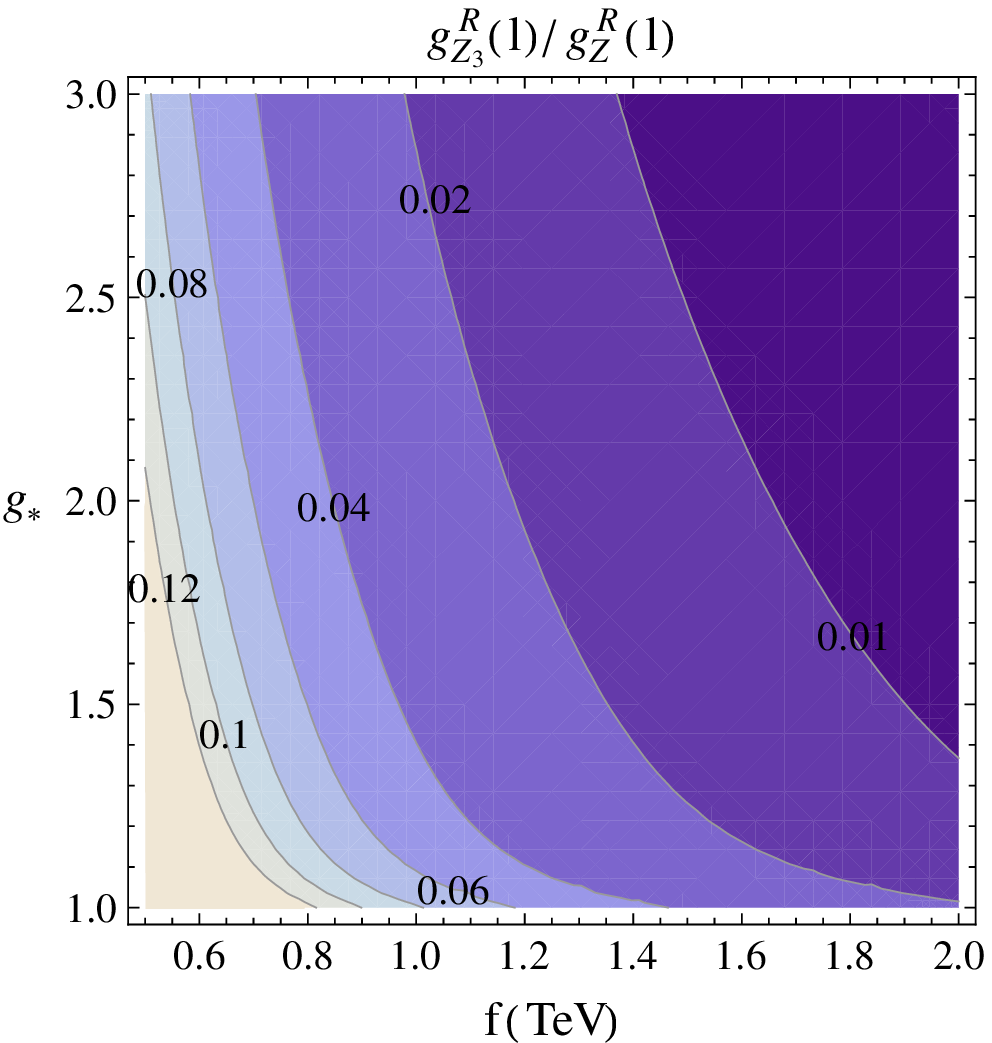,width=7.5cm}}
\end{picture}
\end{center}
\vskip 5.cm \caption{Ratio between the $Z_3$  light-fermion couplings and the $Z$ ones.
Top-left: for $g_{Z_3}^L(l)/g_{Z}^L(l)$. Top-right: for $g_{Z_3}^L(u)/g_{Z}^L(u)$.
Bottom-left: for $g_{Z_3}^L(d)/g_{Z}^L(d)$. Bottom-right: for $g_{Z_3}^R(l)/g_{Z}^R(l)$.}
\label{fig:Z3-femions}
\end{figure}

\begin{figure}[!ht]
\begin{center}
\vspace{-.8cm}
\unitlength1.0cm
\begin{picture}(7,10)
\put(-4.3,2.7){\epsfig{file=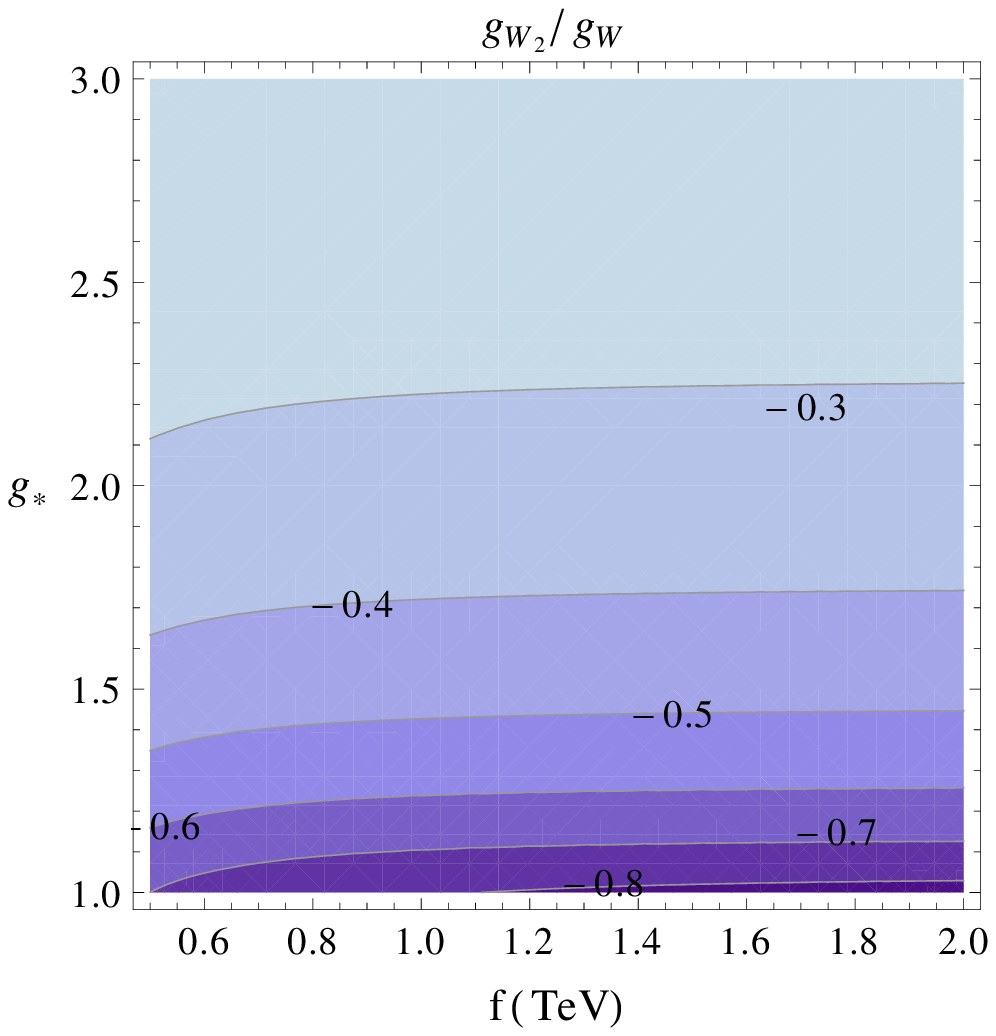,width=7.5cm}}
\put(3.5,2.7){\epsfig{file=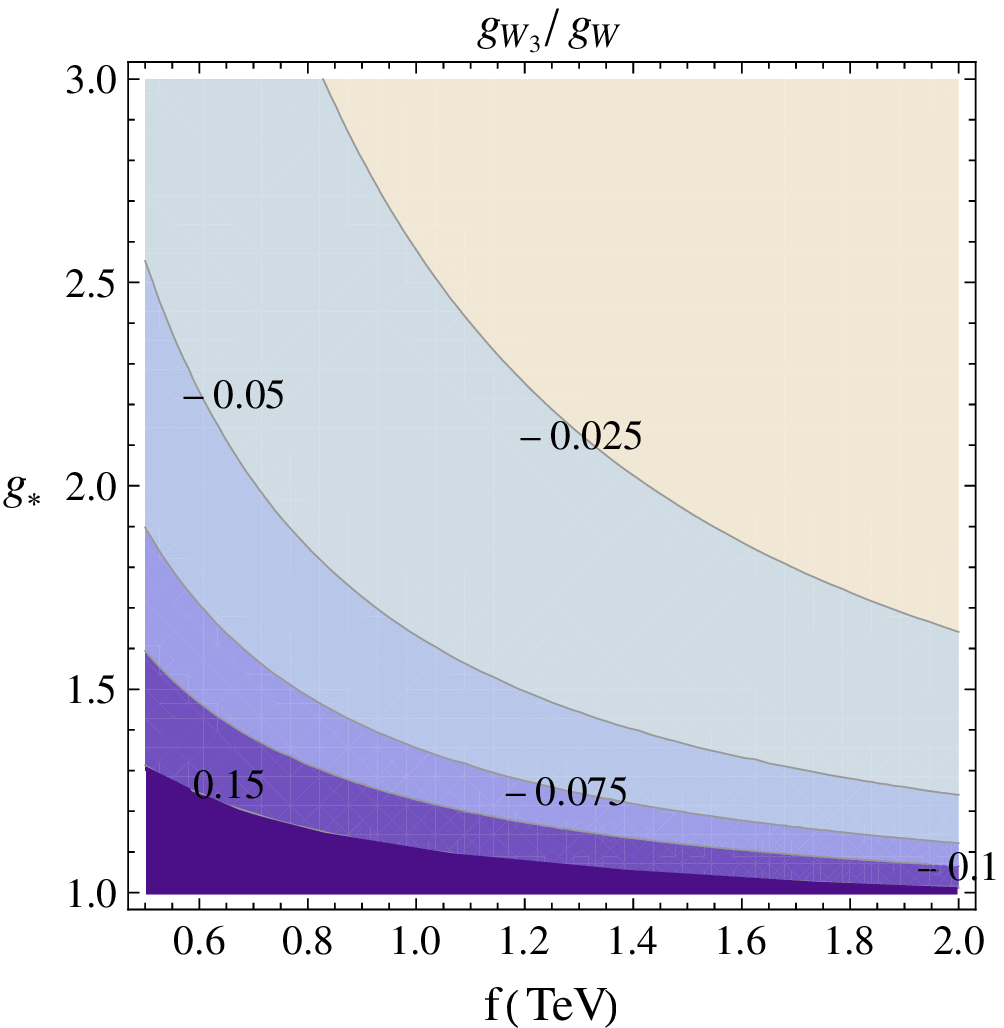,width=7.5cm}}
\end{picture}
\end{center}
\vskip -3cm \caption{Ratio between the $W_2$ and $W_3$ light-fermion couplings and the $W$ ones.
Left: $g_{W_2}/g_{W}$.
Right: $g_{W_3}/g_{W}$.}
\label{fig:W-fermions}
\end{figure}

\begin{figure}[!ht]
\begin{center}
\vspace{1.8cm}
\unitlength1.0cm
\begin{picture}(7,10)
\put(-4.3,2.7){\epsfig{file=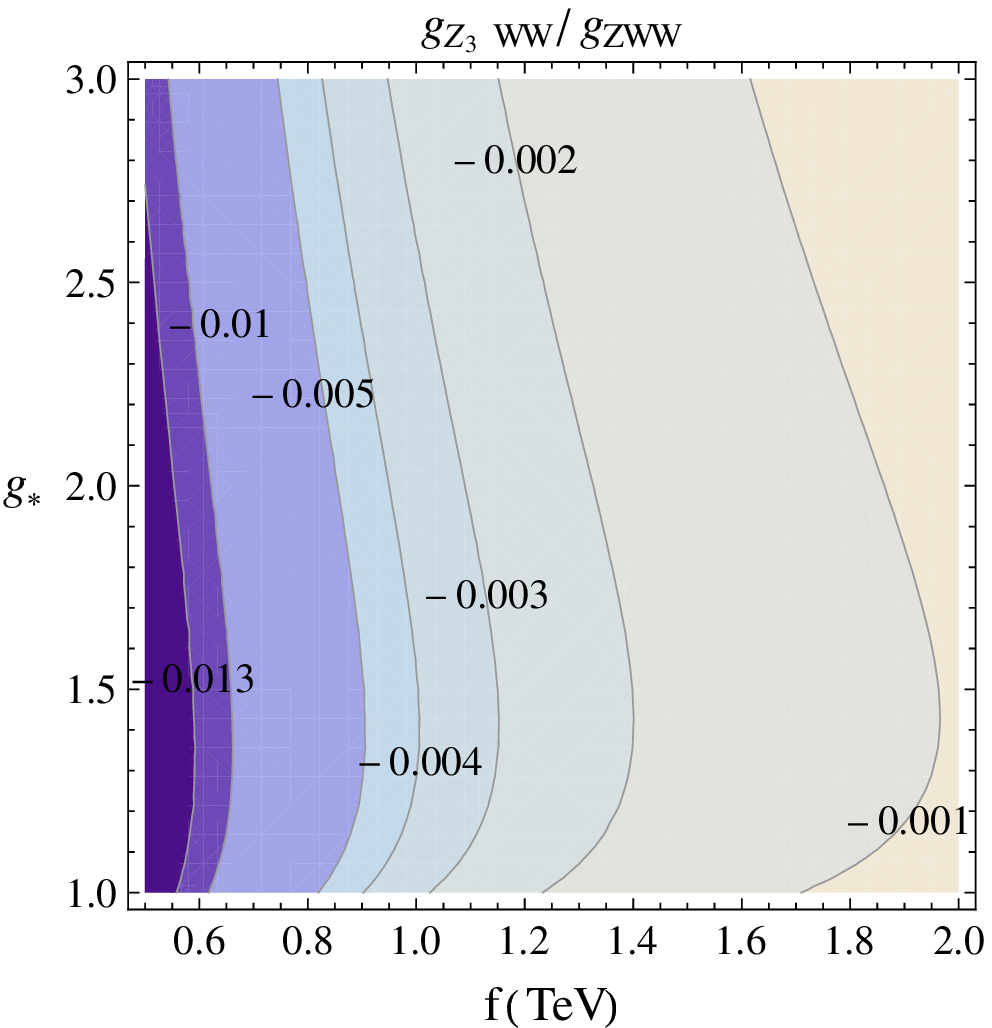,width=7.5cm}}
\put(3.5,2.7){\epsfig{file=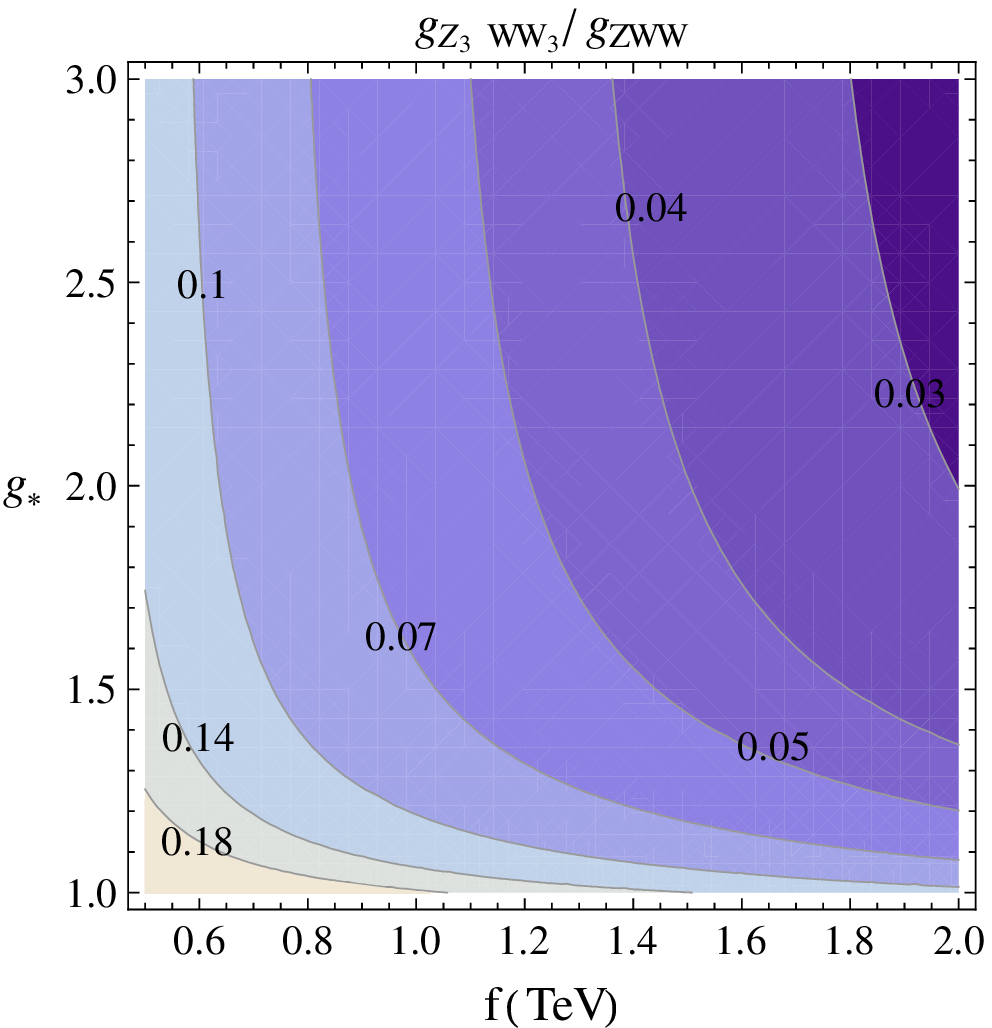,width=7.5cm}}
\end{picture}
\end{center}
\vskip -3cm \caption{Ratio between tri-linear couplings in the 4DCHM and the SM ones.
Left: $g_{Z_3WW}/g_{ZWW}$.
Right: $g_{Z_3WW_3}/g_{ZWW}$.}
\label{fig:VVV}
\end{figure}

Before starting to consider the di-boson results for the 4DCHM, in order to give an idea of the fermion couplings of the new spin-1 resonances, we plot  the ratios between the $Z_{2,3}$ ($W_{2,3}$) light-fermion
couplings and the $Z$  ($W$) ones,  as  functions of the parameters of the 4DCHM in the gauge sector: $f$ and $g_*$. We show only these four bosons because  they are the most relevant ones for the processes we will consider. This is done in  Figs.~\ref{fig:Z2-femions}--\ref{fig:Z3-femions} for the neutral gauge bosons $Z_2$ and 
$Z_3$, respectively
(here, we only show the ratio for the right-handed lepton couplings since the results for the 
corresponding $u$- and $d$-quarks are comparable), and in Fig.~\ref{fig:W-fermions} for the charged gauge bosons $W_2$
and $W_3$. Similar plots for the case of the tri-linear gauge boson couplings (not involving a photon) can be found in Fig.~\ref{fig:VVV}, limited to the two cases 
$g_{Z_3WW}$ and $g_{Z_3WW_3}$, which are those
entering some resonant diagrams in neutral and charged current di-boson production and decay, respectively, that are relevant to our analysis. 
Additional (numerical) values for both the gauge boson to fermion as well as 
tri-linear couplings pertaining to processes (\ref{eq:processWW})--(\ref{eq:processWZ})  are found in Appendix A, for two benchmarks
in the 4DCHM parameter space which will be studied in detail later on. 
  
%%%%%%%%%%%%%%%%%%%%%%%%%%%%%%%%%%%%%%%%%%%%%%%%%%%
%%%%%% SECTION RESULTS           %%%%%%%%%%%%%%%%%%
%%%%%%%%%%%%%%%%%%%%%%%%%%%%%%%%%%%%%%%%%%%%%%%%%%%

\section{Results}
\label{sec:results}

We study in this section the phenomenology of processes (\ref{eq:processWW})--(\ref{eq:processWZ}), 
from the point of view of both their production and decay dynamics. Before doing so though, we briefly describe
the numerical tools used. For more details about the latter, we refer the reader to Refs.~\cite{Accomando:2012yg,DeCurtis:2012cn}.

\begin{figure}[!t]\label{fig:diagrams}
\begin{center}
\hspace*{-5.0truecm}
\includegraphics[scale=0.8]{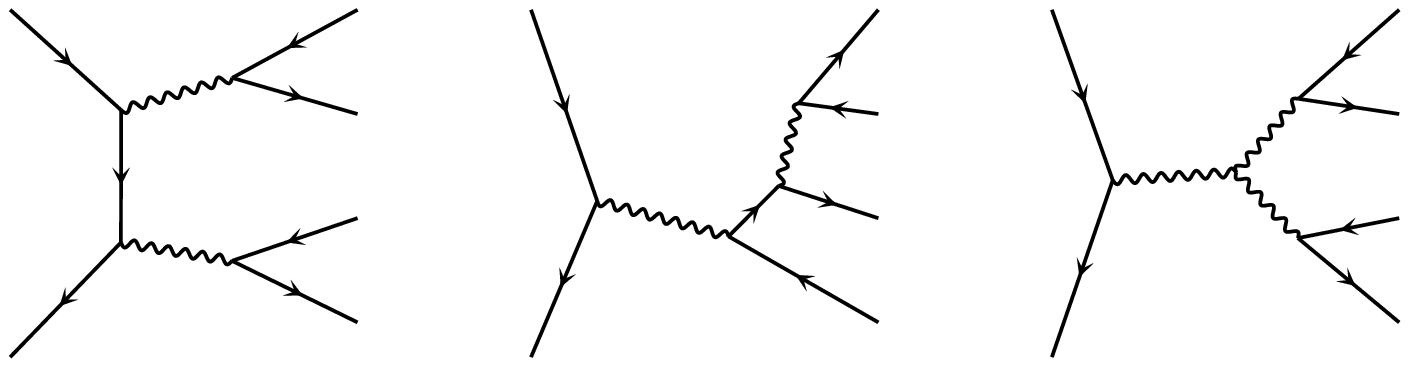}
\end{center}
\vspace*{-19.0truecm}
\caption{Topologies of Feynman diagrams necessary to compute the subprocesses in eqs.~(\ref{eq:processWW})--(\ref{eq:processWZ}).} 
%\vspace*{2.0truecm}
\label{fig:topol}
\end{figure}

\subsection{Calculation}
\label{subsec:calculus}

The numerical results obtained in the previous section for the 4DCHM spectrum generation and tests against experimental data were based on two
codes, one exploiting Mathematica and the other exploiting the LanHEP/CalcHEP environment \cite{Semenov:1996es,Semenov:2010qt,Pukhov:2004ca,Belyaev:2012qa}, cross-checked against each other 
where overlapping\footnote{These modules have been described in detail in Ref.~\cite{Barducci:2012kk} so we do not dwell on them here.
Further, they will be made available shortly on the High Energy Physics Data-Base (HEPMDB) ~\cite{Brooijmans:2012yi}: see
https://hepmdb.soton.ac.uk/.}.

The codes exploited for our study of the LHC signatures
are based on helicity amplitudes, defined through either the PHACT module
\cite{Ballestrero:1999md} or the HELAS subroutines~\cite{Murayama:1992gi}, the latter assembled by means of 
MadGraph~\cite{Stelzer:1994ta}. For both processes (\ref{eq:processWW})--(\ref{eq:processWZ}) we have 
tested the Matrix Elements (MEs) via two independent implementations and it has been verified that both satisfy
standard gauge invariance tests. The MEs account for all off-shellness effects
of the particles involved and were constructed starting from the topologies in Fig.~\ref{fig:topol}, 
wherein the wavy lines refer to any possible gauge bosons in the 4DCHM and the arrow lines to the fermions\footnote{Note that the contribution due to the Higgs process
$pp(gg)\to h\to$  $W^+W^-\to$ $e^+\nu_e$ $ \mu^-\bar\nu_\mu~+~{\rm{c.c.}}$ $\to$ $e^\pm\mu^\mp 
E^{T}_{\rm miss}$ to the $2l$ signature is negligible \cite{Accomando:2012yg}.}.

Two different phase space implementations were also adopted, an `ad-hoc one' (eventually used for event generation) and a `blind one' 
based on RAMBO \cite{Kleiss:1985gy}, again checked one against the other. 
VEGAS~\cite{Lepage:1977sw} was eventually used for the multi-dimensional numerical integrations.

The Parton Distribution Functions (PDFs) used 
were CTEQ5L~\cite{Lai:1999wy}, with factorization/renormalization
scale set to $Q=\mu=\sqrt{\hat{s}}$. Initial state quarks have been taken
as massless, just like the final state leptons and neutrinos. As for the gauge boson sector, its implementation has
been described previously.

it is useful  to introduce also a few kinematic observables that will be used in the remainder
to define acceptance and selection criteria of the two final states in 
processes (\ref{eq:processWW})--(\ref{eq:processWZ}):

\begin{itemize}
\item $\eta_i=-\log(\tan\frac{\theta_i}2)$ is the pseudo-rapidity of a particle,
\item $P^T_{i[j]}=\sqrt{(P_i^x[+P_j^x])^2+(P_i^y[+P_j^y])^2}$ is the transverse momentum of a particle $i$[of a pair of particles $ij$],
\item $p^T_M={\rm max}(P^T_i,P^T_j[,P^T_k])$ is the maximum amongst the transverse momenta of the two[three]  particles,  
\item $P^T_M={\rm max}(P^T_{ij}[,P^T_{ik},P^T_{jk}])$  is the maximum amongst the transverse momenta of all possible pairs of particles,
\item $M_{ij[k]}=\sqrt{(P_i+P_j[+P_k])^\mu(P_i+P_j[+P_k])_\mu}$ is the invariant mass of a pair[tern] of particles, 
\item $M^T_{ij[k]}=\sqrt{(P_i^0+P_j^0[+P_k^0])^2-(P_i^x+P_j^x[+P_k^x])^2-(P_i^y+P_j^y[+P_k^y])^2}$ 
is the transverse mass of a pair[tern] of particles, 
\item $\theta_{i[j]}$ is the angle between the beam axis and a particle[between two particles],
\item $\cos\phi_{ij}^T=\frac{P_i^xP_j^x+P_i^yP_j^y}{P^T_i P_j^T}$ is the cosine of the relative (azimuthal) angle between two particles 
in the plane transverse to the beam,
\item $E^{T}_{\rm miss}=\sqrt{(P_i^x+P_j^x[+P_k^x])^2+(P_i^y+P_j^y[+P_k^y])^2}$ is the missing transverse energy (due to the neutrino 
escaping detection).
\end{itemize}
\noindent
Here, the square brackets are introduced to extend the definition of the observables from the case of the $2l$ to
the $3l$ signature, as the indices $i,j[,k]$ run over the two[three] visible particles in the final state of process
(\ref{eq:processWW})[(\ref{eq:processWZ})], each with momentum $P_i.P_j[,P_k]$.

In contrast to the case of DY processes
dealt with in Ref.~\cite{Barducci:2012kk}, we found no sensitivity here of either of the di-boson channels to charge asymmetries
 (like, e.g., the 
forward-backward one), so that we will not dwell on these observables here. 

Further notice that, in the following, we will use a 10 GeV bin for the forthcoming differential distributions due to the fact that the
 experimental 
resolution (in mass and transverse momentum) is of order 1\% at 1--2~TeV for the electron, while for the muon the rate is like 5\%. 

Finally, as intimated, notice that the benchmark points in the 4DCHM parameter space used here are taken from \cite{Barducci:2012kk},
where their complete parameter listing is given. Namely the (a)--(f) benchmarks are defined in Tab. 20 of Ref. \cite{Barducci:2012kk} 
whereas the {\it colored} ones are given in Tab. 19 of the same paper.

\subsection{The $2l$ signature}

The acceptance and selection cuts that maximize the sensitivity to process (\ref{eq:processWW}) have been defined
in Ref.~\cite{Accomando:2012yg} (where they were referred to as {\it So} cuts):
$$
|\eta_{e,\mu}|<2,\qquad
p^T_{e,\mu}>20~{\rm GeV},\qquad
E^{T}_{\rm miss}>50~{\rm GeV},
$$
\begin{equation}\label{eq:So}
M_{e\mu}>180~{\rm GeV},\qquad 
p_T^M<300~{\rm GeV},\qquad
\cos\phi_{e\mu}^T<-0.9,\qquad
\cos\theta_{e\mu}<0.5.
\end{equation}

\begin{figure}[!t]
\begin{center}
\vspace{-.8cm}
\unitlength1.0cm
\begin{picture}(7,10)
\put(-4.3,2.7){\epsfig{file=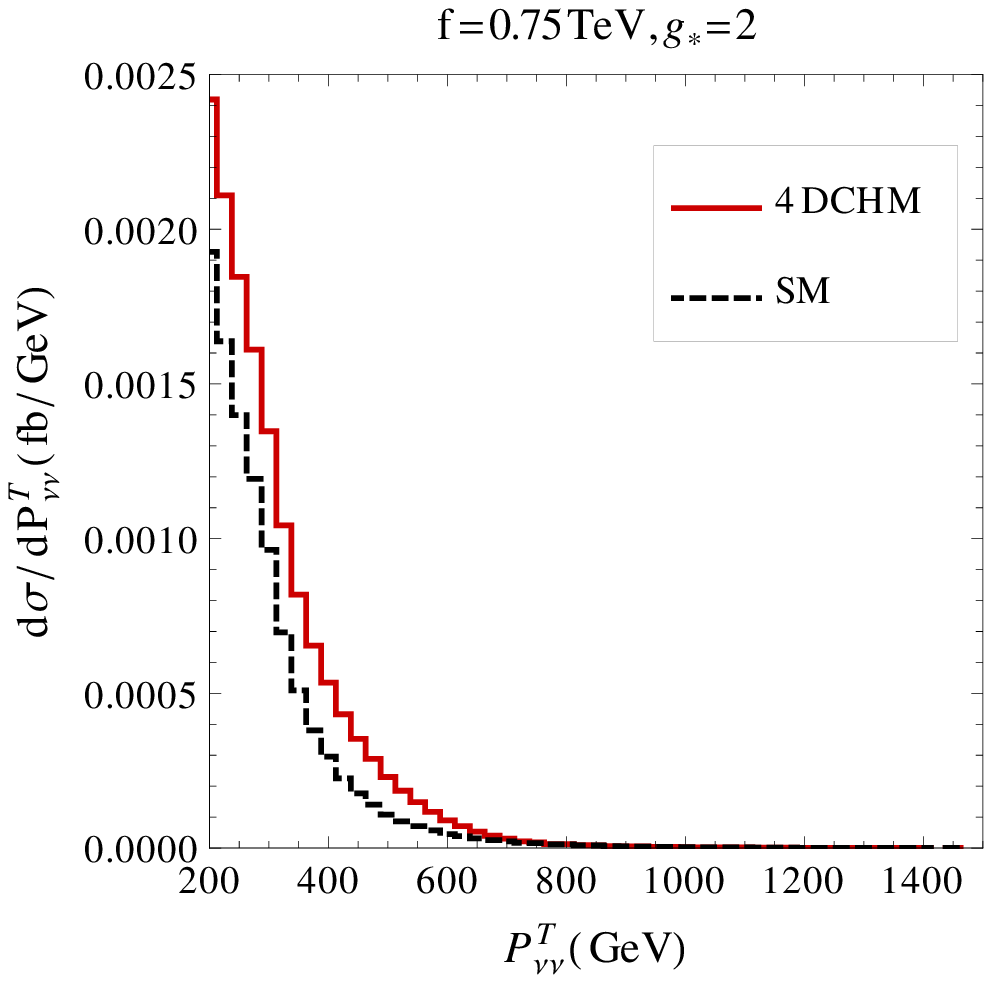,width=7.15cm}}
\put(3.5,2.7){\epsfig{file=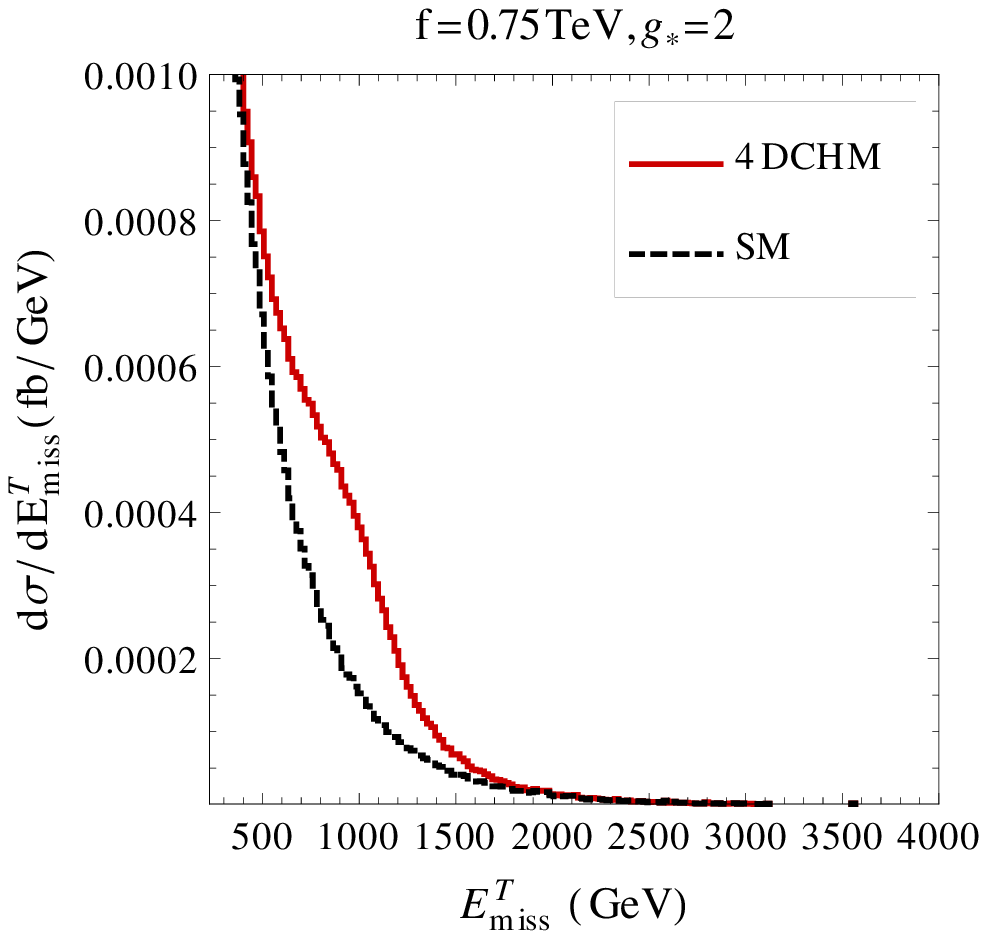,width=7.5cm}}
\put(-4.3,-5){\epsfig{file=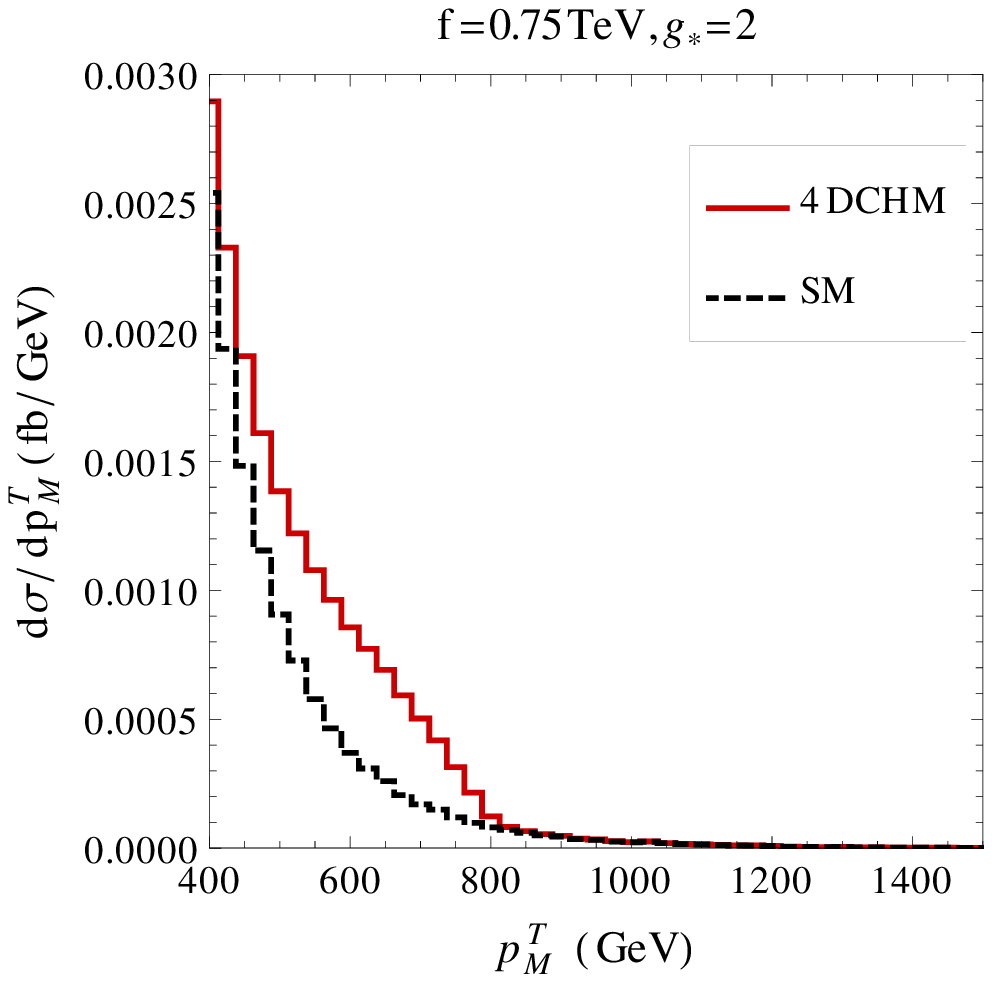,width=7.15cm}}
\put(3.5,-5){\epsfig{file=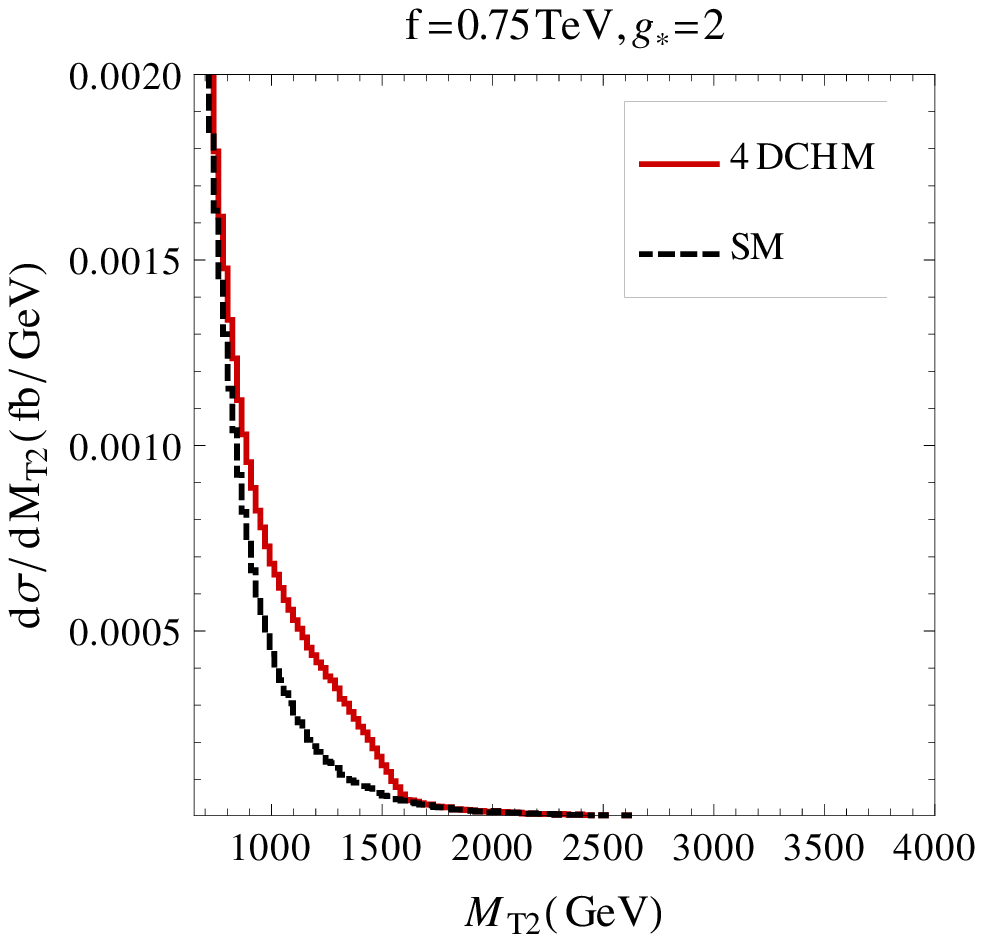,width=7.5cm}}
\end{picture}
\end{center}
\vskip 4.cm \caption{Differential cross sections pertaining to the di-boson process in (\ref{eq:processWW}) {at the 14~TeV LHC}.
 Here, $p^T_M$ and $E^T_{\rm miss}$ are defined in \protect{Subsect.~\ref{subsec:calculus}},
$P^T_{\nu\nu}$ is the transverse momentum of the two neutrinos in the plane transverse to the beam 
whereas $M_{T2}$ is the transverse mass as defined in \protect{Ref.~\cite{AA:2012ks}}. {\it So} cuts are applied. 
The red-solid curve represents the full 4DCHM whilst the black-dashed one refers to the SM. The benchmark (a) $f=0.75$ TeV
and $g_*=2$ of \protect{\cite{Barducci:2012kk}} is adopted here.
}
\label{fig:distr_75-2}
\end{figure}

\begin{figure}[!ht]
\begin{center}
\vspace{-.8cm}
\unitlength1.0cm
\begin{picture}(7,10)
\put(-4.3,2.7){\epsfig{file=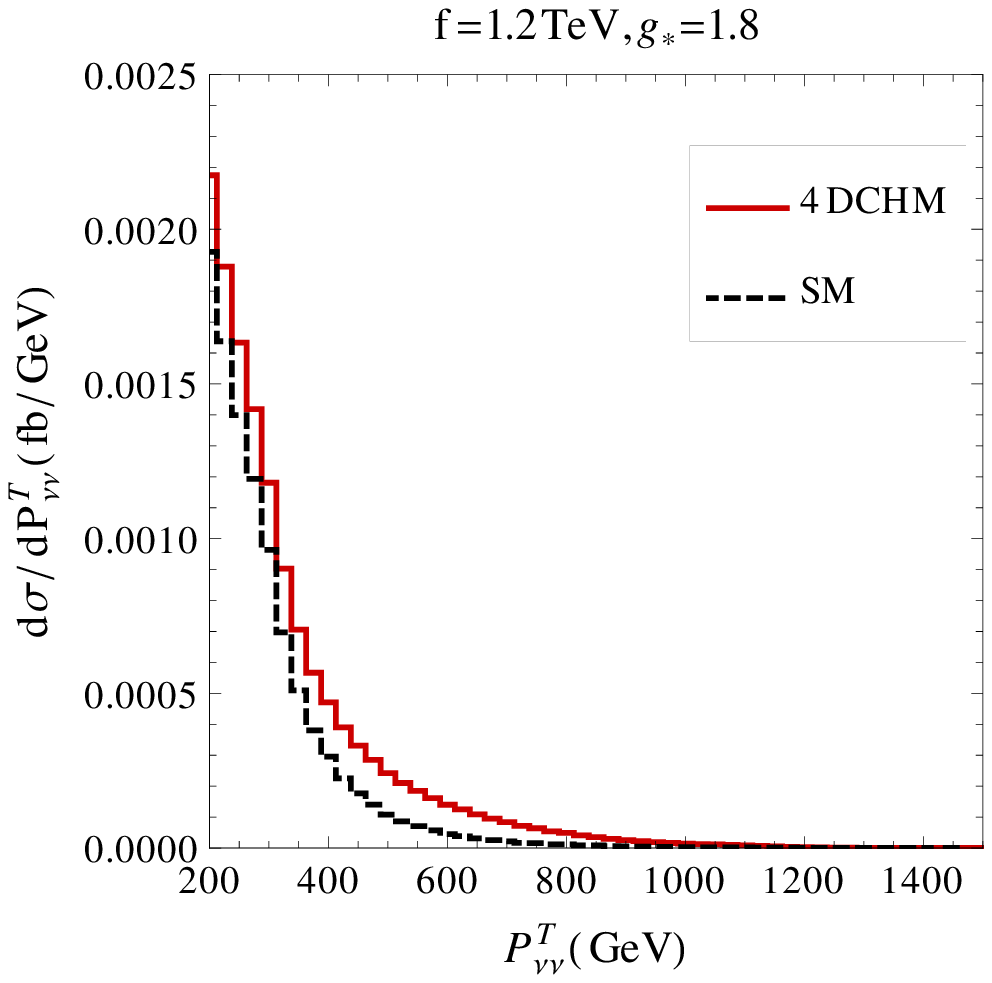,width=7.15cm}}
\put(3.5,2.7){\epsfig{file=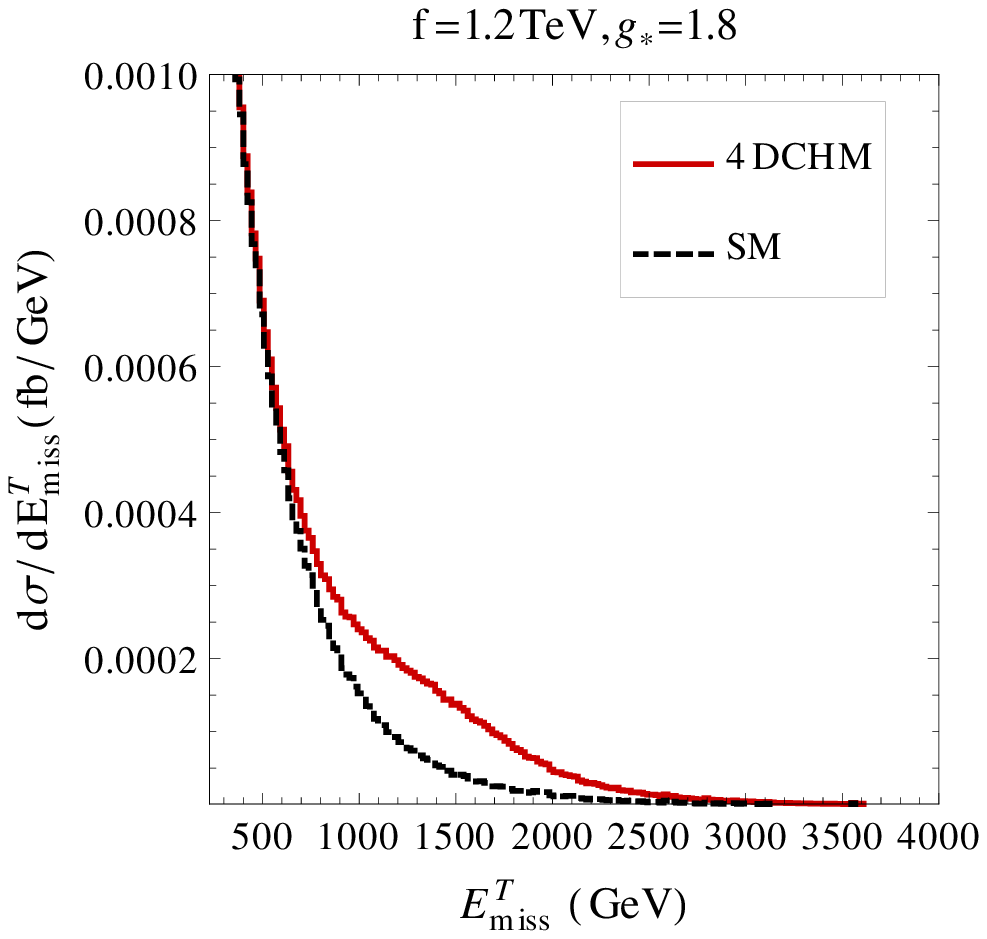,width=7.5cm}}
\put(-4.3,-5){\epsfig{file=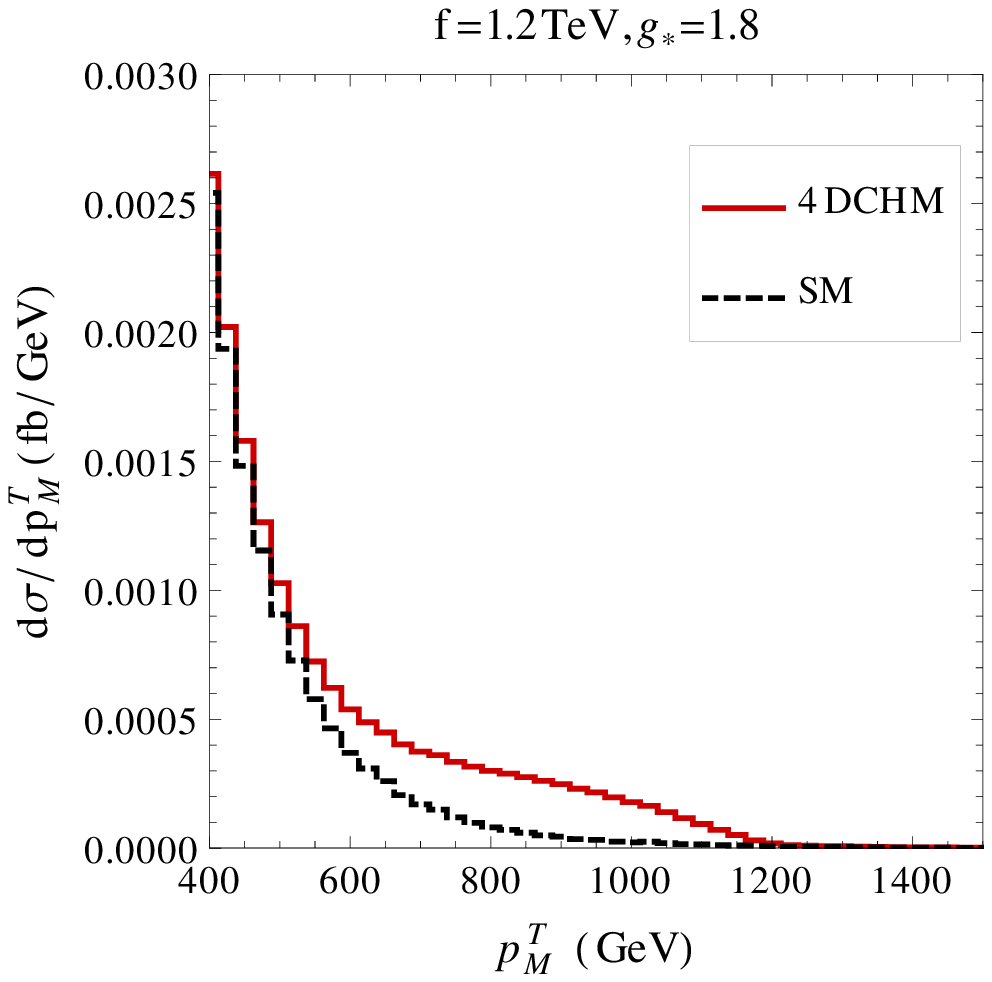,width=7.15cm}}
\put(3.5,-5){\epsfig{file=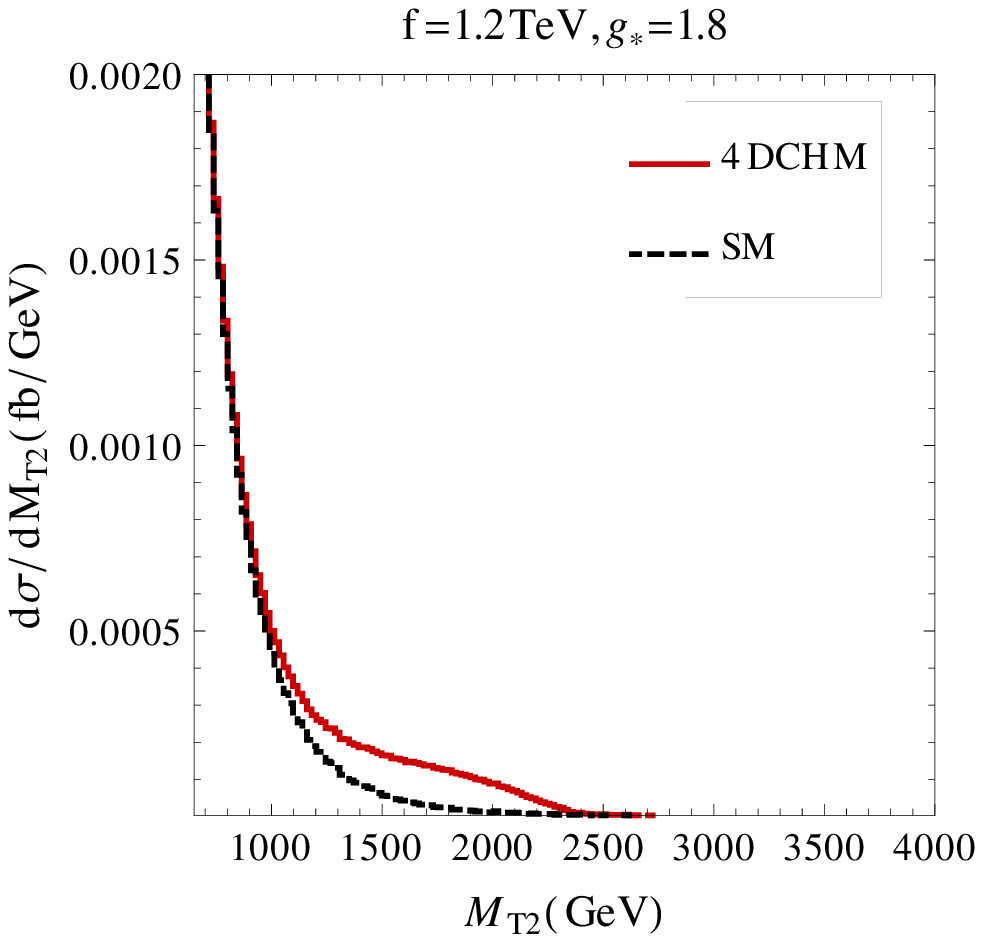,width=7.5cm}}
\end{picture}
\end{center}
\vskip 4.cm \caption{Same as in \protect{Fig.~\ref{fig:distr_75-2}} for the benchmark (f) $f=1.2$ TeV
and $g_*=1.8$ of \protect{\cite{Barducci:2012kk}}.
}
\label{fig:distr_12-18}
\end{figure}

In Figs.~\ref{fig:distr_75-2}--\ref{fig:distr_12-18} we show some relevant observables for two of the
4DCHM  benchmarks defined in \cite{Barducci:2012kk}, in particular (a) and (f), respectively. 
The fact that is not possible to detect all the final state particles, and
in particular that there are two invisible neutrinos, makes it very hard to achieve a clear identification of the 
intermediate vector bosons. However, 
the effects of extra neutral gauge boson resonances appear as an excess of events in some energy measure below 
the value corresponding to the new gauge boson mass. 
(Unfortunately,  there is no observable that allows one to have a signature of the new charged bosons involved in the process.) 
As in the 4DCHM the $Z_{2,3}$ bosons
are close in mass, it is  impossible to separate them, and,  between the two, it is the $Z_3$ state the one
with typically largest cross section, \cite{Barducci:2012kk}. Moreover, the $Z_5$  state is very weakly coupled to 
the SM light-fermions and very heavy (see Appendix A), consequently, it is essentially invisible. Hence, the results in these figures
essentially  highlight
the $Z_3$ mass as the end point of the excess region extending to the left of $M_{Z_3}$ in case of energy measures 
($E^T_{\rm miss}$ and $M_{T2}$) or to the left of $M_{Z_3}/2$ for the transverse 
momentum measures ($P^T_{\nu\nu}$ and $p^T_M$). Although not shown, the pattern emerging for all the other benchmarks defined in
Ref.~\cite{Barducci:2012kk} is similar to the one illustrated here. These distributions were obtained for a LHC energy of
14~TeV, 
however, they are rather similar {in shape (but not in magnitude) to those at 7 and 8~TeV}.

In Tabs.~\ref{Tab:CS_multi} and~\ref{Tab:CS_12-18} we present the cross sections of 
channel (\ref{eq:processWW}), at 8 and 14 TeV (not at 7 TeV, as corresponding rates are
generally poor)  for each of the twelve benchmarks of \cite{Barducci:2012kk} (notice that `(f)' and `red' are actually the same) after 
the cuts given in eq.~(\ref{eq:So}). No further  selection is adopted here, as the 4DCHM curves consistently sit above the SM
ones in  Figs.~\ref{fig:distr_75-2}--\ref{fig:distr_12-18} over the entire kinematic ranges of the 
observables considered. Hereafter, we define as signal ($S$) the difference between the 
total ($T$) 4DCHM result and the SM one, the latter thereby constituting the background (B). In order to extract from these rates the 
statistical significance,
$\sigma$, it  is enough to multiply the last column for $\sqrt{L\epsilon}$, where $L$ is the luminosity in fb$^{-1}$ and $\epsilon$ is the 
efficiency
to tag the final state. If we consider a luminosity of, e.g., 25~fb$^{-1}$ at the 14~TeV LHC and an overall tagging 
{efficiency of} {55\%}
(as obtained  in Ref.~\cite{Accomando:2012yg}), we have, e.g., for the cyan benchmark, that 
 the minimal $S/\sqrt{B}$~(in $\sqrt{\rm fb}$ units)
necessary to have a statistical significance of 5 (discovery) is 0.35~$\sqrt{\rm fb}$ whereas for exclusion (i.e., statistical 
significance 2) we
need at last 0.14~$\sqrt{\rm fb}$. 
In general, it is clear  that, at 14 TeV, over the generic 4DCHM parameter space, the $Z_{3}$ boson  can easily be extracted
(i.e., for all benchmarks (a)--(f) considered, Tab.~\ref{Tab:CS_multi}), so long that its width is small enough since (consider the 
$colored$ benchmarks now,
illustrating the effect for $f=1.2$ TeV and $g_*=1.8$, Tab.~\ref{Tab:CS_12-18}),
with increasing values of this quantity, the scope of the LHC in this channel diminishes significantly (as illustrated in 
Ref.~\cite{Barducci:2012kk}, the smaller the mass
of the composite fermions the larger the widths of the gauge bosons in general, as the latter are allowed to decay in the former). 
Prospects at 8 TeV are
instead rather negative, certainly for detection and most probably for exclusion too. Finally, independently of the energy and luminosity 
of the LHC, none of the
other 4DCHM neutral or charged gauge boson resonances is accessible via channel (\ref{eq:processWW}).
\begin{table}[!t]
\begin{center}
\begin{tabular}{||c|c|c||c|c||}
\hline \hline
~~~~&(f) (TeV)& $g_*$& $S=T-B$~(fb) & $S/\sqrt{B}$~($\sqrt{\rm fb}$)\\
\hline 
\hline
(a)&0.75&2   &1.8 (0.19)   &  2.6 (0.4)\\
(b)&0.8 &2.5 &0.22 (0.024) &  0.32 (0.05)\\
$(c)$&1   &2   &0.36 (0.039) &  0.52 (0.081)\\
(d)&1   &2.5 &0.07 (0.0046)&  0.10 (0.0096)\\
(e)&1.1 &1.8 &0.42 (0.046) &  0.60 (0.096)\\
(f)&1.2 &1.8 &0.24 (0.022) &  0.34 (0.046)\\
%&0.5&4&0.0032&0.0045\\
%&1.5&1.33&0.36&0.47\\
\hline\hline
\end{tabular}
\end{center}
\caption{Cross sections for process (\ref{eq:processWW}) in the 4DCHM at the 14(8)~TeV LHC, using \textit{So} cuts.
The SM background is 0.49(0.23)~fb.   The benchmarks (a)--(f) of \protect{\cite{Barducci:2012kk}} are adopted here.
%ricordarsi di moltiplicare x2 le CS per includere ee+mumu
}
\label{Tab:CS_multi}
\end{table} 
\begin{table}[!ht]
\begin{center}
\begin{tabular}{||c||c|c||}
\hline \hline
$f=1.2$ TeV, $g_*=1.8$&$S=T-B$~(fb) & $S/\sqrt{B}$~($\sqrt{\rm fb}$)\\
\hline 
\hline
red    &0.24 (0.022)  & 0.34 (0.046)\\
%red$^*$&0.38 (0.038)  & 0.52\\
green  &0.15 (0.014)  & 0.22 (0.029)\\
cyan   &0.14 (0.013)  & 0.20 (0.027)\\
magenta&0.12 (0.011)  & 0.18 (0.023)\\
black  &0.028 (0.0024)& 0.04 (0.005)\\
yellow &0.012 (0.0010)& 0.016 (0.0021)\\
\hline\hline
\end{tabular}
\end{center}
\caption{Cross sections for process (\ref{eq:processWW}) in the 4DCHM at 14(8)~TeV LHC, using \textit{So} cuts.
The SM background is 0.49(0.23)~fb.   The $colored$ benchmarks of \protect{\cite{Barducci:2012kk}} are adopted here.
%ricordarsi di moltiplicare x2 le CS per includere ee+mumu
}
\label{Tab:CS_12-18}
\end{table} 

\subsection{The $3l$ signature}

Before proceeding with the study of this signature, a subtlety should be noted.
Some of the variables defined in Subsect.~\ref{subsec:calculus} implicitly assume the capability to identify in the final state of process (\ref{eq:processWZ})
the two leptons coming from the neutral current (propagated by the $\gamma,Z,,Z_{2},Z_3,Z_5$ states). In the 
$l=e$ and $l'=\mu$ (or vice versa) case this is trivial, since the pair of leptons
with identical flavor are necessarily those emerging from such a current. In the case $l=l'=e$ or $\mu$ the identification
is in principle ambiguous (incidentally, this requires an anti-symmetrization of the diagrams stemming from the topologies in Fig. \ref{fig:topol},
which we have done, according to Pauli-Dirac statistics). However, in Ref.~\cite{DeCurtis:2012cn}, an efficient method was devised to overcome this problem, by noting that $P_M^T$ is
generally the one induced by the pair of partons emerging from the $\gamma,Z,Z_{2},Z_3,Z_5$ current, so that this enables use to enforce the same cuts onto
the final state of the  process (\ref{eq:processWZ}), irrespectively of the actual $l,l'$ combination being generated.

The acceptance and selection criteria that maximize the sensitivity to process (\ref{eq:processWZ}) have been introduced 
in Ref.~\cite{DeCurtis:2012cn} (where they were called {\it C2} cuts):
$$
|\eta_{l^\pm,l^{'+},l^{'-}}|<2,\qquad
P^T_{l^\pm,l^{'+},l^{'-}}>20~{\rm GeV},\qquad 
E^{T}_{\rm miss}>50~{\rm GeV},
$$
$$
M_{ l^\pm l^{'+},~l^\pm l^{'-},~l^{'+}l^{'-}}>20~{\rm GeV},~~
p^T_M>150~{\rm GeV},~~
\cos\phi_{l^{'+}l^{'-}}^T<-0.5,~~
\cos\theta_{l^\pm l^{'+},l^\pm l^{'-},l^{'+}l^{'-}}<0.9
$$
\begin{equation}\label{eq:C2}
P^T_{l^\pm l^{'+},~l^\pm l^{'-},~l^{'+}l^{'-}}>150~{\rm GeV},\qquad\qquad 
M_{l^\pm l^{'+} l^{'-}}>0.9 ~M_{W_2}.
\end{equation}
The last cut, which depends on a 4DCHM parameter, unlike the others, is actually justified by the fact that the sequence of {\it C2} restrictions is enforced in steps,
so that, after the first seven cuts have been implemented (first two rows of eq.~(\ref{eq:C2})), the $W_2$ resonance clearly emerges above the background, as illustrated in Ref.~\cite{DeCurtis:2012cn} (albeit for another 
model)\footnote{Alternatively,
following Ref.~\cite{Barducci:2012kk}, the extraction of a value for $M_{W_2}$ is always possible in the DY
channel.}.   

In this part of the paper, we intend to show that process (\ref{eq:processWZ}) at the LHC 
(again, we will take $\sqrt s=14$ TeV in the plots for illustration purposes) can act
as an effective means to extract part of the mass spectrum of the gauge sector of the 4DCHM that is not accessible elsewhere.  
 In order to accomplish this, it is crucial the fact that process (\ref{eq:processWZ}) affords one with the possibility to reconstruct
 the missing longitudinal momentum of the neutrino, as
also described in \cite{DeCurtis:2012cn}.  Therefore, alongside $M_{l^{'+}l^{'-}}$ (which is sensitive to the neutral gauge boson 
resonances), we can also plot the
reconstructed Centre-of-Mass (CM) energy at the partonic level, $\sqrt{\hat s}=E^R_{\rm cm}$ (which is sensitive to the charged gauge
 boson resonances)\footnote{Any sensitivity to the latter in the henceforth reconstructed invariant mass distribution of the charged $l^\pm \nu_l$ current 
is lost, owing to the fact
that our aforementioned procedure of computing the longitudinal neutrino momentum selects only the $W$ component, 
i.e., that of the SM \cite{DeCurtis:2012cn}.}. 

\begin{figure}[!htbp]
\begin{center}
% \vspace{.0cm}
\unitlength1.0cm
\begin{picture}(7,10)
\put(-4.3,2.7){\epsfig{file=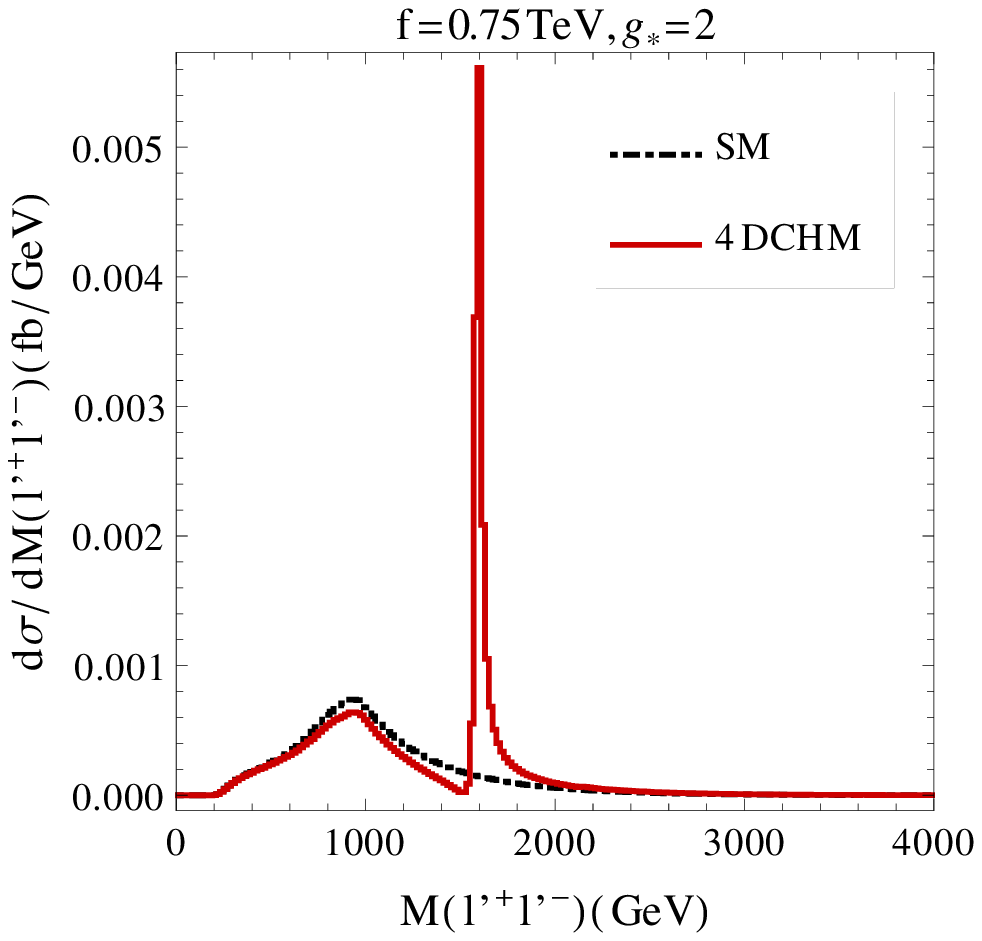,width=7.5cm}}
\put(3.5,2.7){\epsfig{file=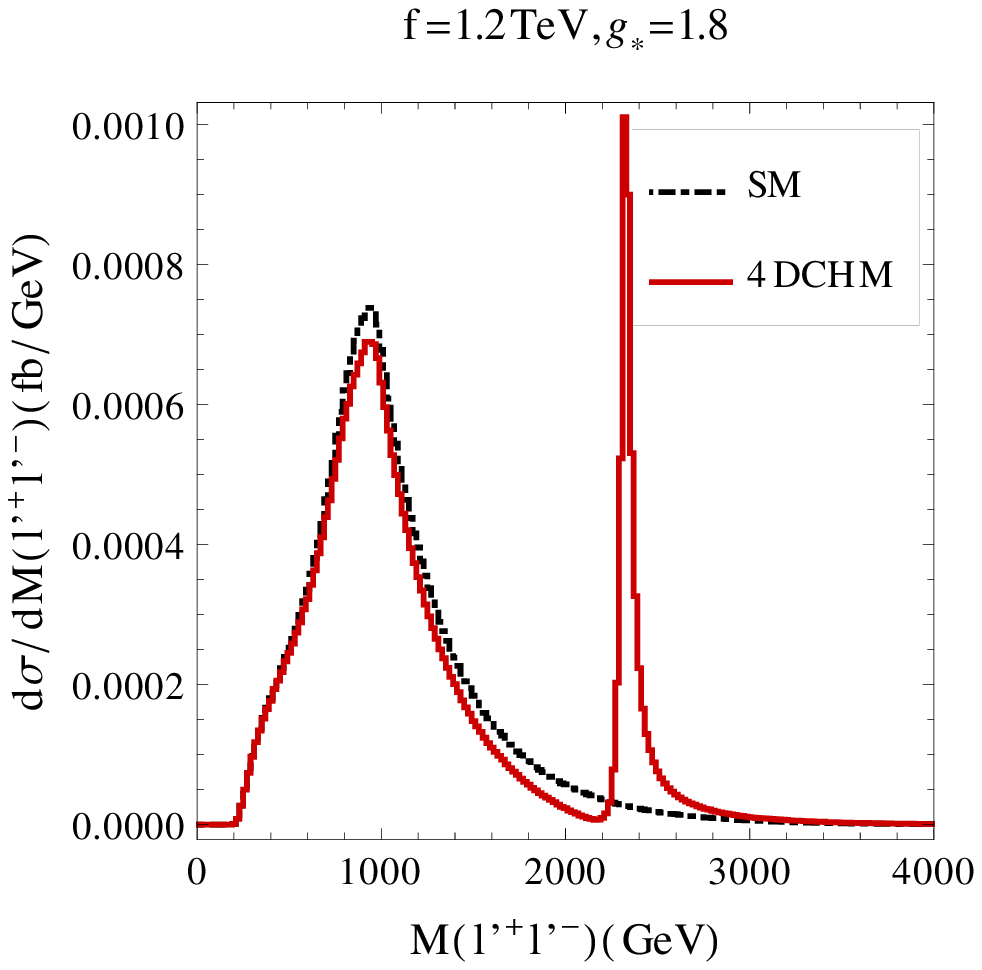,width=7.5cm}}
\put(-4.3,-5){\epsfig{file=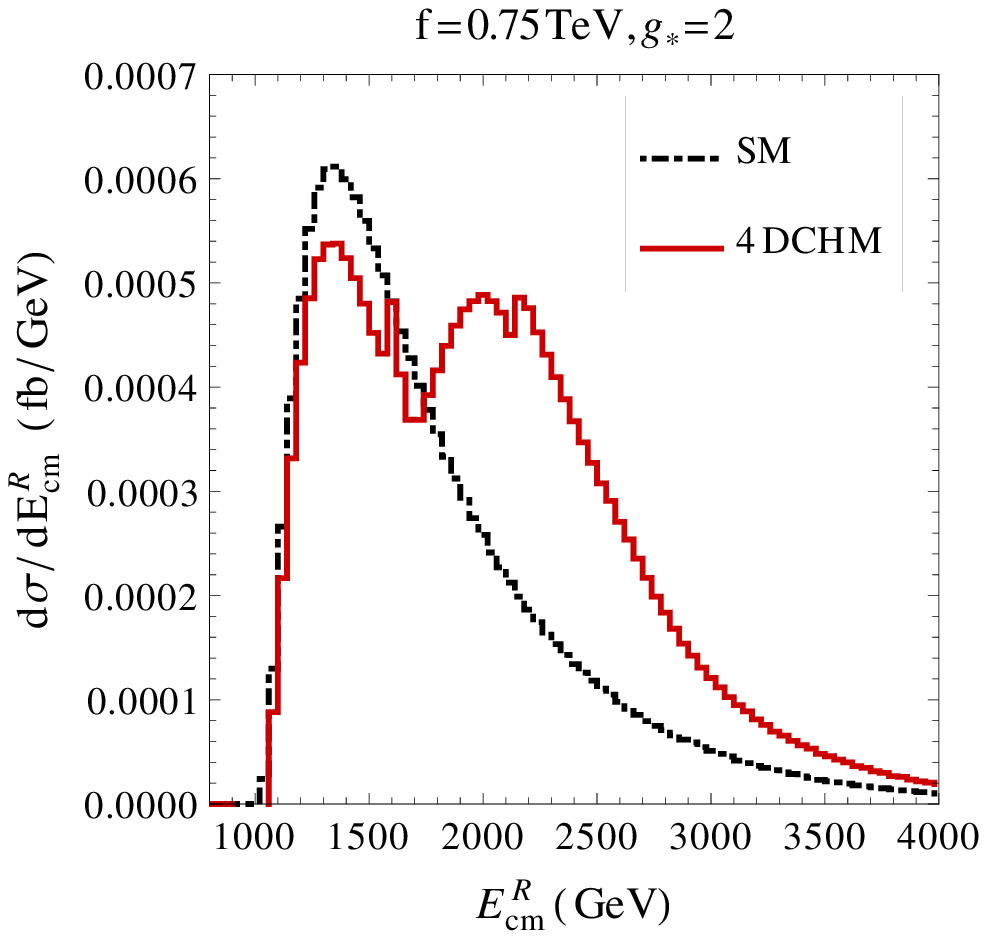,width=7.5cm}}
\put(3.5,-5){\epsfig{file=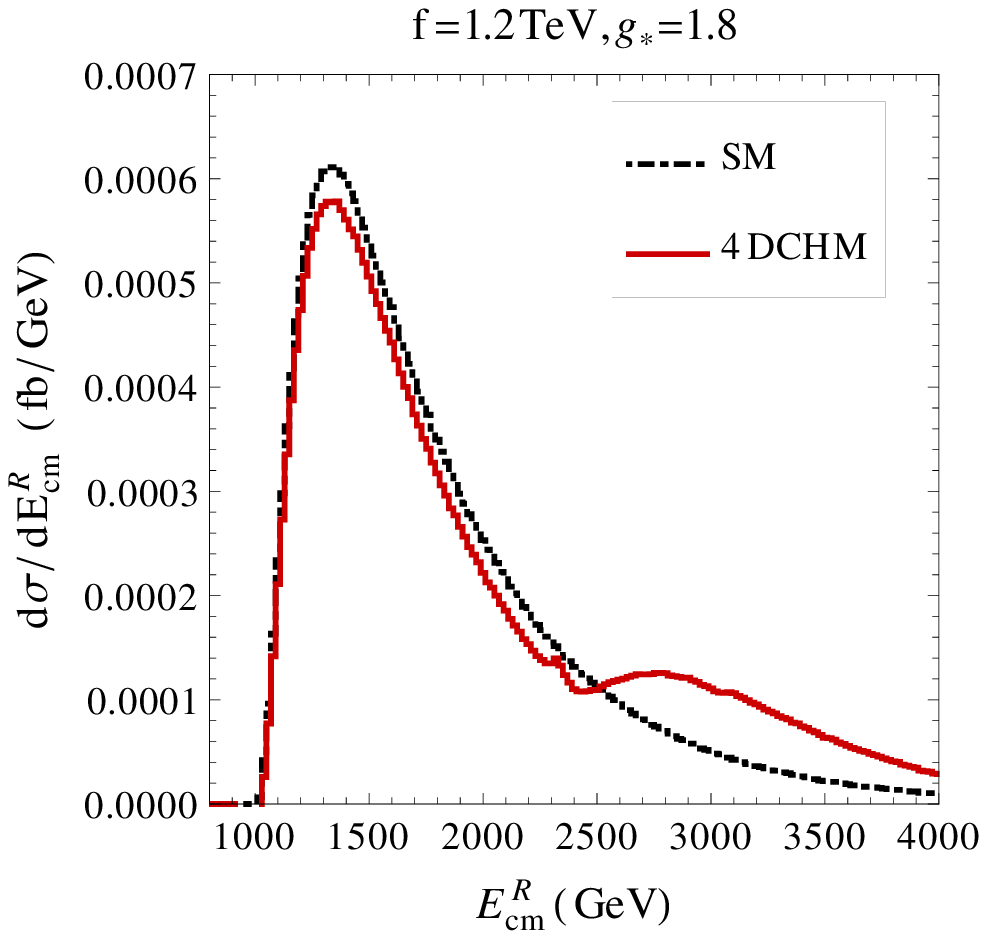,width=7.5cm}}
\end{picture}
\end{center}
\vskip 4.cm \caption{
Differential cross sections pertaining to the di-boson process in (\ref{eq:processWZ}).
 Here, $M_{l^{'+}l^{'-}}$   (top)  and $E^R_{\rm cm}$ (bottom) are defined in  \protect{Subsect.~\ref{subsec:calculus}} and in
the running text, respectively. {\it C2} cuts are applied. 
The red-solid curve represents the full 4DCHM whilst the black-dashed one refers to the SM. The benchmark (a) $f=0.75$ TeV
and $g_*=2$[(f) $f=1.2$ TeV
and $g_*=1.8$] of \protect{\cite{Barducci:2012kk}} is adopted here on the left[right] hand side.}
\label{fig:distr_C1_M56_M456}
\end{figure}

Again, the benchmarks chosen here from Ref.~\cite{Barducci:2012kk}, representative of a situation occurring over the whole of
the 4DCHM parameter space, are the (a) and (f) ones. Fig.~\ref{fig:distr_C1_M56_M456} plots the aforementioned two kinematic variables, 
from which we recognize, again, the presence of the $Z_2$ (partially) and $Z_3$ (mainly) resonances in the  $M_{l^{'+}l^{'-}}$ spectrum 
as well as that of the $W_3$ resonance, in fact for the first time,
in the $E_{\rm cm}^R$ one.  The latter occurrence is peculiar to process
(\ref{eq:processWZ}), as it did occur neither in the DY modes investigated in  \cite{Barducci:2012kk} nor in channel (\ref{eq:processWW})
 studied here. In contrast,
the $Z_5$, again, does not emerge over the background in the invariant mass of the di-lepton pair because of, needless to say, its small 
couplings and large mass. Finally the $W_2$, whilst evident in the reconstructed CM energy at partonic level, is clearly mimicked by the 
SM background, in view of the last cut
in eq.~(\ref{eq:C2}), which renders the signal and background very similar.

Before proceeding to establish the significance of the $Z_{2,3}$ and $W_3$ mass peaks, one peculiar feature of process 
(\ref{eq:processWZ})
should be noted, i.e., unlike the case of (\ref{eq:processWW}), it is characterized by  a, at times, large negative interference between 
the diagrams involving the
new gauge bosons of the 4DCHM and those involving the SM ones, induced by the signs of the fermion-gauge boson as well as tri-linear couplings of the 4DCHM with
respect to those of the SM (see \cite{Barducci:2012kk}), which can onset precisely where the masses of the new gauge bosons are. In fact,
 such an
effect is ultimately responsible for the generic smallness of the cross sections for process  (\ref{eq:processWZ}) in comparison 
to 
reaction  (\ref{eq:processWW}). Hence, for some combinations of 4DCHM inputs, the actual signal is a depletion of the expected SM rate,
in the relevant
mass region. This is manifest in Tabs.~\ref{Tab:banch}--\ref{Tab:banch2}, which show the signal and background rates 
 integrated over the $M_{l^{'+}l^{l'-}}$ range between 2 and 
3 TeV (except for benchmark (a) for which we integrate from 1.5 to 2 TeV), i.e., the $Z_{2,3}$ peak region,
and  over the $E_{cm}^R$  range between 2.5 TeV and 4 TeV (except for benchmark (a) for which we integrate from 2 to 4 TeV), i.e., 
the $W_3$ peak region, respectively. In the same Tables, upon assuming an overall tagging {efficiency of} 
{50\%}~\cite{DeCurtis:2012cn}, we 
present the significance $\sigma$ of the signal for 1000 fb$^{-1}$ of luminosity as well as its minimal value required to claim detection
 ($5\sigma$).
(Notice that in the calculation of the significance, for the cases where the signal appears as a depletion rather than an excess, as it 
can be 
for the $W_3$ peak but not the $Z_{2,3}$ one, we have taken the modulus
of the signal rates.) While extraction of the signal is possible in a few instances with standard luminosities expected at the LHC 
running at 14 TeV, in 
others this requires much larger data samples, probably obtainable only at the currently considered Super-LHC stage \cite{Gianotti:2002xx}
(i.e., a tenfold increase in instantaneous luminosity for the 14 TeV LHC). The need for high luminosity is clearly the stronger the 
larger the 
gauge boson masses  and/or widths involved.

\begin{table}[!ht]
\begin{center}
\begin{tabular}{||c|c|c|c|c||}
\hline \hline
~~~~ & $S$~(fb)&$S/\sqrt{B}$~($\sqrt{\rm fb}$)& $\sigma$&$L_{m}$~(fb$^{-1}$)\\
\hline 
\hline
(a)&1.1   &2.2  &65  &5.9\\
(b)&0.067 &0.23 &5.2 &924\\
$(c)$&0.25  &0.85 &19  &69\\
(d)&0.0061&0.021&0.47& NA\\
(e)&0.45  &1.6  &35  &20\\
(f)&0.3  &1.0  &23  &47\\
\hline\hline
\end{tabular}
%\quad
\begin{tabular}{||c|c|c|c|c||}
\hline \hline
 & $S$~(fb)&$S/\sqrt{B}$~($\sqrt{\rm fb}$)& $\sigma$&$L_{m}$~(fb$^{-1}$)\\
\hline \hline
red    & 0.3 & 1.0  &23 & 47\\
green  &0.18  & 0.62 &14 &128\\
cyan   &0.17  & 0.58 &13 &148\\
magenta&0.15  & 0.5 &11 &207\\
black  &0.029 & 0.099&2.2&5165\\
yellow &0.014 & 0.048&1.1& NA\\
\hline\hline
\end{tabular}
\end{center}
\caption{Cross sections for process (\ref{eq:processWZ}) in the 4DCHM at the 14~TeV LHC,
using \textit{C2} cuts supplemented by an additional selection 
 around the $Z_{2,3}$ mass (see the text). 
 The benchmarks (a)--(f) (left sub-table) as well as the $colored$ ones (right sub-table) of \protect{\cite{Barducci:2012kk}} are adopted 
here.
The SM background is always  0.085~fb, except for benchmark (a) which yields
0.27~fb.
The statistical significance $\sigma$ is computed with 
a luminosity of 1000 fb$^{-1}$ and
the tagging efficiency mentioned in the text. $L_{m}$ is the minimal luminosity needed to discover the $Z_{2,3}$ peak. The label ``NA'' is related to luminosity values which are not accessible at present colliders as well as future proposed prototypes.
}
\label{Tab:banch}
\end{table}
\begin{table}[!b]
\begin{center}
\begin{tabular}{||c|c|c|c|c||}
\hline \hline
~~~~ & $S$~(fb)&$S/\sqrt{B}$~($\sqrt{\rm fb}$)& $\sigma$&$L_{m}$~(fb$^{-1}$)\\
\hline 
\hline
(a)&0.81   &1.1   &24 &43\\
%(b)&0.031  &0.064 &1.4&1.2$\times10^{4}$\\
(b)&0.031  &0.064 &1.4& NA \\
(c)&0.19   &0.39  &8.8&322\\
%(d)&$-0.0081$&$-0.016$&0.36  &1.9$\times10^{5}$\\
(d)&$-0.008$&$-0.016$&0.36  & NA \\
(e)&0.38   &0.77  &17 &86\\
(f)&0.24   &0.48  &11 &206\\
\hline\hline
\end{tabular}
%\quad
\begin{tabular}{||c|c|c|c|c||}
\hline \hline
 & $S$~(fb)&$S/\sqrt{B}$~($\sqrt{\rm fb}$)& $\sigma$&$L_{m}$~(fb$^{-1}$)\\
\hline \hline
red    &0.24   &0.48   &11 &206\\
green  &0.13   &0.26   &5.9&718\\
cyan   &0.12   &0.24   &5.5&826\\
magenta&0.097  &0.2   &4.5&1234\\
%black  &$-0.0042$&$-0.0086$&0.19  &6.9$\times10^{5}$\\
%yellow &$-0.009$&$-0.018$ &0.41  &1.5$\times10^{5}$\\
black  &$-0.004$&$-0.009$&0.19  & NA \\
yellow &$-0.009$&$-0.018$ &0.41  & NA \\
\hline\hline
\end{tabular}
\end{center}
\caption{Cross sections for process (\ref{eq:processWZ}) in the 4DCHM at the 14~TeV LHC,
using \textit{C2} cuts supplemented by an additional selection 
 around the $W_{3}$ mass (see the text). 
 The benchmarks (a)--(f) (left sub-table) as well as the $colored$ ones (right sub-table) of \protect{\cite{Barducci:2012kk}} are adopted here.
The SM background is always  0.24~fb, except for benchmark (a) which yields
0.57~fb.
The statistical significance $\sigma$ is computed with 
a luminosity of 1000 fb$^{-1}$ and
the tagging efficiency mentioned in the text. $L_{m}$ is the minimal luminosity needed to discover the $W_3$ peak. The label ``NA'' is related to luminosity values which are not accessible at present colliders as well as future proposed prototypes.
}
\label{Tab:banch2}
\end{table}

\begin{figure}[!htbp]
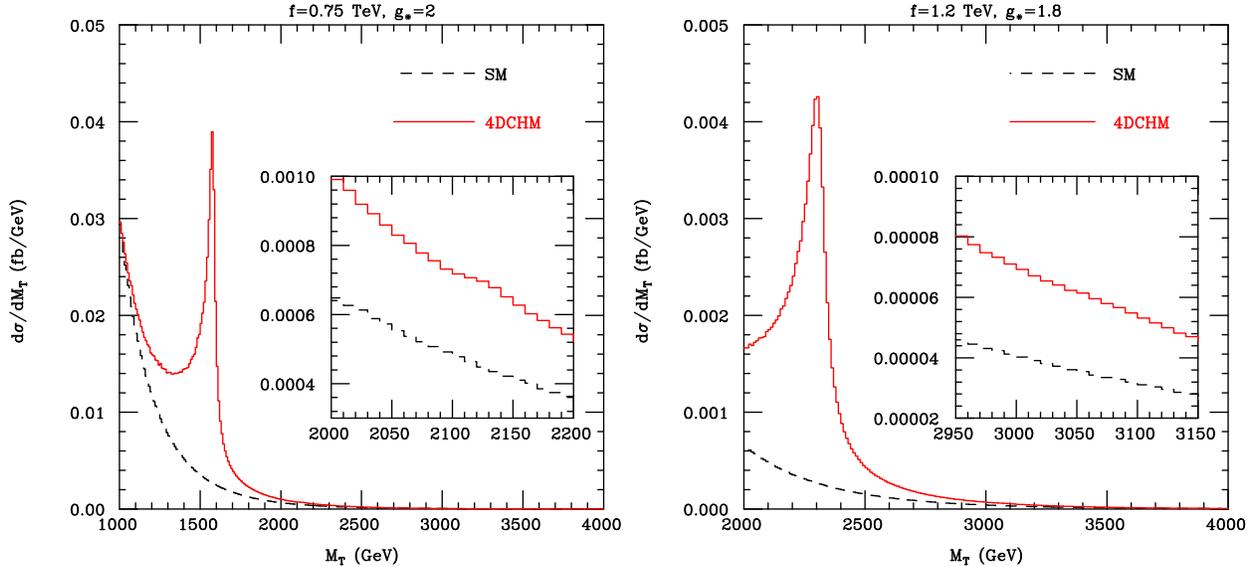

\begin{center}
\unitlength1.0cm
\begin{picture}(7,10)
\put(-4.8,2.7){\epsfig{file=DY_CC-Spettrof075g2_cuts-MT.ps,width=7.5cm,angle=90}}
\put(3.5,2.7){\epsfig{file=DY_CC-Spettrof12g18_cuts-MT.ps,width=7.5cm,angle=90}}
\end{picture}
\end{center}
\vskip -3.5cm \caption{
Differential cross sections pertaining to the charged DY process $pp(q\bar q')\to W^\pm\to l^+\nu_l~+~{\rm c.c.}\to l^\pm E_{\rm miss}^T$.
 Here, $M_T$ is the transverse mass defined in the running text. 
The red-solid curve represents the full 4DCHM whilst the black-dashed one refers to the SM. The benchmark (a) $f=0.75$ TeV
and $g_*=2$[(f) $f=1.2$ TeV
and $g_*=1.8$] of \protect{\cite{Barducci:2012kk}} is adopted here on the left[right] hand side.
The cuts enforced are 
$p^T_l>20$ GeV, $|\eta_l|<2.5$ and 
$M_{T}>1.0$ TeV (left) or $M_{T}>2.0$ TeV (right).}
\label{fig:DY}
\end{figure}

Of particular relevance here is to notice that, while accessing the $Z_2$ and $Z_3$ resonances is always generally possible 
through the neutral current DY process, this is not the typical case for the $W_3$ resonance via the charged current DY channel, see Ref.~\cite{Barducci:2012kk}. This is made explicit by consulting Fig.~\ref{fig:DY}, where we plot the spectrum in the transverse mass of
the $l^\pm E_{\rm miss}^T$ final state emerging in the latter case, i.e., of
$M_T\equiv\sqrt{(E^T_l+E^T_{\rm miss})^2
             -(p^x_{l}+p^x_{\rm miss})^2
             -(p^y_{l}+p^y_{\rm miss})^2}$,
where $E^T$ represents missing energy/momentum (as we consider the electron and muon massless) in the plane transverse to the beam 
and $p_{x,y}$ are the two components therein (assuming that the proton beams are directed along the $z$ axis). Again, we refer to 
benchmarks (a) and (f) of Ref.~\cite{Barducci:2012kk}, for which a Jacobian shape reminiscent of the $W_3$ mass
is expected to appear around the $M_T$ values of
2123 and 3056 GeV, respectively (see Tab.~21 in \cite{Barducci:2012kk}). This is clearly not the case and in fact it turns out 
that it is the contribution due to the interference between the lighter $W_2$ peak (occurring at 1581 and 2312 GeV, respectively,
see again Tab.~21 in \cite{Barducci:2012kk})  and the SM
contributions that overwhelms the $W_3$ peak. Contrast this figure with the two bottom plots in Fig.~\ref{fig:distr_C1_M56_M456}.

Finally, if one computes the event rates for process (\ref{eq:processWZ}) after the cuts in eq.~(\ref{eq:C2}),
as done in Tabs.~\ref{Tab:CSWZ_multi}--\ref{Tab:CSWZ_12-18} (i.e., without any resonance extraction), the devastating effects of the aforementioned negative interferences are apparent, to the extent that most of the 4DCHM benchmarks (hence, parameter space) 
become inaccessible at the standard LHC and the Super-LHC would become the only available
option. Hence, it is clear that a high resolution sampling in various kinematic distributions is of fundamental importance to establish a signal in this channel at the
LHC and this is clearly impossible at the 7 and 8 TeV energy stages, given the limited data samples collected therein.

\begin{table}[!t]
\begin{center}
\begin{tabular}{||c|c|c||c|c||}
\hline \hline
~~~~ &(f) (TeV)& $g_*$& $S=T-B$~(fb) & $S/\sqrt{B}$~($\sqrt{\rm fb}$)\\
\hline 
\hline
(a)&0.75&2   &0.78   & 0.48\\
(b)&0.8 &2.5 &$-7.8\times10^{-2}$ & $-4.8\times10^{-2}$\\
(c)&1   &2   &$7.4\times10^{-2}$ & $3.6\times10^{-2}$\\
(d)&1   &2.5 &$-6.7\times10^{-2}$&  $-4.1\times10^{-2}$\\
(e)&1.1 &1.8 &0.26 & 0.16\\
(f)&1.2 &1.8 &0.11 &  $6.9\times10^{-2}$\\
%&0.5&4&0.0032&0.0045\\
%&1.5&1.33&0.36&0.47\\
\hline\hline
\end{tabular}
\end{center}
\caption{Cross sections for process (\ref{eq:processWZ}) in the 4DCHM at the 14~TeV LHC,
using \textit{C2} cuts. 
 The benchmarks (a)--(f) of \protect{\cite{Barducci:2012kk}} are adopted here.
The SM background is 2.6~fb.}
\label{Tab:CSWZ_multi}
\end{table} 
\begin{table}[!ht]
\begin{center}
\begin{tabular}{||c||c|c||}
\hline \hline
$f=1.2$ TeV, $g_*=1.8$&$S=T-B$~(fb) & $S/\sqrt{B}$~($\sqrt{\rm fb}$)\\
\hline 
\hline
red    &0.11& $6.9\times10^{-2}$\\
%red$^*$&0.15&$9.4\times10^{-2}$\\
green  &$-1.4\times10^{-2}$&$-8.7 \times10^{-3}$\\
cyan   &$-2.6\times10^{-2}$& $-1.6\times10^{-2}$\\
magenta&$-5.2\times10^{-2}$ &$-3.3\times10^{-2}$ \\
black  &$-0.16$ &$-9.7\times10^{-2}$\\
yellow &$-0.12$ & $-7.3\times10^{-2}$\\
\hline\hline
\end{tabular}
\end{center}
\caption{Cross sections for process (\ref{eq:processWZ}) in the 4DCHM at the 14~TeV LHC,
using \textit{C2} cuts. 
 The colored benchmarks of \protect{\cite{Barducci:2012kk}} are adopted here.
The SM background is 2.6~fb.}
\label{Tab:CSWZ_12-18}
\end{table} 

\section{Conclusions}
\label{sec:summa}

In summary, in the
%perennial
recurring study of
%prominent
leptonic signatures neatly emerging from the hadronic noise of the LHC, 
if contrasted to the potential of DY processes, the scope of both charged and mixed di-boson production in enabling one to test
the gauge sector of the 4DCHM is promising, despite the unavoidable lower cross sections. In fact, the benefit of supplementing standard DY analyses
with di-boson ones is twofold. On the one hand, the latter, unlike the former, are anyhow sensitive to tri-linear gauge boson vertices, which are required to
be measured in order to uniquely pinpoint the underlying gauge structure. On the other hand, the di-boson modes,
other than confirming the presence of the lightest gauge boson resonances already accessible in DY channels (and possibly measuring their widths), also offer the chance to extract the heaviest of the new charged gauge boson resonances of the 4DCHM, which would escape searches in DY mode. Unfortunately, the heaviest of the new neutral gauge boson resonances eludes the reach of both DY and di-boson processes. A residual possibility to access it would then be via gauge boson decays into hadrons
in general and specifically via third generation quarks \cite{Ken:2012pp}, 
which mix with the additional heavy fermions of the 4DCHM, so that the intervening coupling strength between 4DCHM gauge bosons and SM top- and bottom-quark 
 is higher than in the case of leptons (and light quarks as well).   

\section*{Acknowledgments}
\noindent
We thank Alexander Belyaev, Diego Becciolini and Elena Accomando for help with various implementations of process and most useful discussions.
DB, AB and SM are financed in part through the NExT Institute. LF thanks Fondazione Della Riccia for financial support.  DB and GMP thank the Galileo Galilei Institute for Theoretical Physics for the hospitality and the INFN for partial support during the completion of this work.
GMP is supported by the German Research Foundation DFG through Grant No.\ STO876/2-1 and by BMBF Grant No.\ 05H09ODEGMP.

%%%%%%%%%%%%%%%%%%%%%%%%%%%%%%%%%%%%%%%%%%%
%%%% APPENDIX BENCHMARK POINTS %%%%%%%%%%%
%%%%%%%%%%%%%%%%%%%%%%%%%%%%%%%%%%%%%%%%%%
\appendix
\section{Benchmark points}
In this additional Section,
we give the numerical values for masses, widths, couplings to the SM light-fermions (introduced in eqs.~(\ref{LCC}), (\ref{LNC})) of the $Z$ and all $Z'$s as well as the $W$ and all $W'$'s, 
and also for
 the tri-linear couplings among all the spin-1 gauge bosons defined by:
\begin{equation}
g_{V'V''V'''}\epsilon_{abc}(\partial_{\mu}A^{'a}_{\nu})(A^{''\mu,b}A^{'''\nu,c}).
\end{equation}
We will here present these values  for the two benchmark points used throughout the paper, namely (a) and (f).
%%%%%%%%%%%%%%%%%%%%%%%%%%%%%%%%%%%%%%%%%%%
%%%% f075g2         %%%%%%%%%%%
%%%%%%%%%%%%%%%%%%%%%%%%%%%%%%%%%%%%%%%%%%
\subsection{Benchmark point (a): $f=0.75$ TeV and $g_*=2$}
Relevant numerical values can be found in Tabs.~\ref{tab:f075g2-masses}--\ref{tab:f075g2-VVV}.
\begin{table}[htb]
\centering
\begin{tabular}{||c|c|c||c|c|c||}
\hline\hline
&$M$ (GeV) &$\Gamma$ (GeV) & & $M$ (GeV) &$\Gamma$ (GeV) \\
\hline\hline
$Z$ & 91.2 &2.4 & $W^{\pm}$ & 80 &2.0\\
%\hline
$Z_2$ & 1549 &28 & $W_2^{\pm}$&1581 &26\\
%\hline
$Z_3$ &1581 & 26& $W_3^{\pm}$& 2123&33\\
%\hline
\cline{4-6}
$Z_5$ &2124 & 34 \\
\cline{1-3}
\end{tabular}
\caption{Masses and widths of the neutral and charged gauge bosons of the 4DCHM for the case $f=0.75 $ TeV and $g_*=2$.}
\label{tab:f075g2-masses}
\end{table}
\begin{table}[htb]
%\captionsetup[subfloat]{labelformat=empty,position=top}
\centering
\begin{tabular}{||c|c|c|c|c||c|c||}
\hline\hline
&$g^{L/R}_{Z_i}(l)$        &$g^L_{Z_i}(\nu)$        &$g^{L/R}_{Z_i}(u)$        &$g^{L/R}_{Z_i}(d)$        &         & $g_{W^{\pm}_i}$        \\
\hline\hline
$Z$        & $-0.20/0.17$        & $0.37$         &$0.26/-0.11$        &$-0.31/0.057$        &$W^{\pm}$        & $-0.46$        \\
%\hline
$Z_2$        &$-0.039/-0.091$        &$-0.052$         &0.009/0.061        &$0.022/-0.030$        &$W_2^{\pm}$ &0.15        \\
%\hline
$Z_3$        & 0.11/0.009        &$-0.10$         &$-0.10/-0.006$        &0.10/0.003       &$W_3^{\pm}$         &0.022        \\
%\hline
\cline{6-7}
$Z_5$        &$0.011/-0.008$        &$-0.019$         &$-0.014/0.006$        &$0.017/-0.003$         \\
\cline{1-5}
\end{tabular}
\caption{Left- and right-handed couplings of the neutral and charged gauge bosons of the 4DCHM to  leptons and the first generation of quarks
for the case $f=0.75 $ TeV and $g_*=2$.}
\label{tab:f075g2-ffV}
\end{table}
%%%%%%%%%%%%%%%%%%%%%%%%%%%%%%%%%%%%%%%%%%%%%v
%%% ACCOPPIAMENTI DI GAUGE
%%%%%%%%%%%%%%%%%%%%%%%%%%%%%%%%%%%%%%%%%%%%%%
\begin{table}[htb]
%\captionsetup[subfloat]{labelformat=empty,position=top}
\centering
\begin{tabular}{||c||c|c|c|c||}
\hline\hline
$Z$         & $W^{+}$         &$W_2^{+}$         &$W_3^{+}$        \\
\hline
\hline
$W^{-}$        &$-0.57$ &        0.004        &        $-8.8\cdot 10^{-6}$        \\
%\hline
$W_2^{-}$        &0.004        &         $-0.55$&        $-0.052$        \\
%\hline
$W_3^{-}$        &$-8.8\cdot 10^{-6}$        &$-0.052$&        $-0.20$        \\
\hline\hline
\end{tabular}
\begin{tabular}{||c||c|c|c|c||}
\hline\hline
$Z_2$         & $W^{+}$         &$W_2^{+}$         &$W_3^{+}$        \\
\hline
\hline
$W^{-}$        &$-0.002$        &        $-0.049$        &0.025        \\
%\hline
$W_2^{-}$        &$-0.049$        &        $-0.13$        &$-0.006$        \\
%\hline
$W_3^{-}$        &0.025        &        $-0.006$&        0.59        \\
\hline\hline
\end{tabular}\\
\vspace{0.1cm}
\begin{tabular}{||c||c|c|c|c||}
\hline\hline
$Z_3$         & $W^{+}$         &$W_2^{+}$         &$W_3^{+}$         \\
\hline
\hline
$W^{-}$        &0.004         &        $-0.63$ &         $-0.047$        \\
%\hline
$W_2^{-}$        &$-0.63$ &        $-1.7$ &         0.031        \\
%\hline
$W_3^{-}$        &        $-0.047$ &        0.031         &         $-1.0$        \\
\hline\hline
\end{tabular}
\begin{tabular}{||c||c|c|c|c||}
\hline\hline
$Z_5$         & $W^{+}$         &$W_2^{+}$         &$W_3^{+}$        \\
\hline
\hline
$W^{-}$        &$-1.9\cdot 10^{-5}$ &         $-0.045$&        $-0.33$         \\
%\hline
$W_2^{-}$        &$-0.045$         &0.032        &$-0.98$        \\
%\hline
$W_3^{-}$        &$-0.33$         &$-0.98$&        $-0.041$        \\
\hline\hline
\end{tabular}
\caption{Tri-linear coupling of the neutral to the charged gauge bosons of the 4DCHM for the case $f=0.75 $ TeV and $g_*=2$.}
\label{tab:f075g2-VVV}
\end{table}
%%%%%%%%%%%%%%%%%%%%%%%%%%%%%%%%%%%%%%%%%%%
%%%% f12g18         %%%%%%%%%%%
%%%%%%%%%%%%%%%%%%%%%%%%%%%%%%%%%%%%%%%%%%
\subsection{Benchmark point (f): $f=1.2 $ TeV and $g_*=1.8$}
In this case, numerical values can be found in Tabs.~\ref{tab:f12g18-masses}--\ref{tab:f12g18-VVV}.
\begin{table}[htb]
\centering
\begin{tabular}{||c|c|c||c|c|c||}
\hline\hline
&$M$ (GeV) &$\Gamma$ (GeV) & & $M$ (GeV) &$\Gamma$ (GeV) \\
\hline\hline
$Z$ &91.2 &2.4 & $W^{\pm}$ & 80 &2.0\\
%\hline
$Z_2$ &2249 &32 & $W_2^{\pm}$&2312 &55\\
%\hline
$Z_3$ &2312 &55 & $W_3^{\pm}$&3056 &54\\
%\hline
\cline{4-6}
$Z_5$ &3056 &54 \\
\cline{1-3}
\end{tabular}
\caption{Masses and widths of the neutral and charged gauge bosons of the 4DCHM for the case $f=1.2 $ TeV and $g_*=1.8$.}
\label{tab:f12g18-masses}
\end{table}
\begin{table}[htb]
%\captionsetup[subfloat]{labelformat=empty,position=top}
\centering
\begin{tabular}{||c|c|c|c|c||c|c||}
\hline\hline
&$g^{L/R}_{Z_i}(l)$        &$g^L_{Z_i}(\nu)$        &$g^{L/R}_{Z_i}(u)$        &$g^{L/R}_{Z_i}(d)$        &         & $g_{W^{\pm}_i}$        \\
\hline\hline
$Z$        & $-0.20/0.17$        &         0.37       &$0.26/-0.11$         &$-0.31/0.057$        &$W^{\pm}$        & $-0.46$        \\
%\hline
$Z_2$        &$-0.049/-0.10$        &        $-0.054$        &0.015/0.069         &$0.020/-0.034$        &$W_2^{\pm}$ & 0.17         \\
%\hline
$Z_3$        & 0.13/0.004        &        $-0.12$        &$-0.12/-0.002$         &0.12/0.001        &$W_3^{\pm}$ &        0.016        \\
%\hline
\cline{6-7}
$Z_5$        &$0.009/-0.006$        &        $-0.014$        &$-0.010/0.004$         &$0.012/-0.002$ \\
\cline{1-5}
\end{tabular}
\caption{Left- and right-handed couplings of the neutral and charged gauge bosons of the 4DCHM to leptons and the first generation of quarks
for the case $f=1.2 $ TeV and $g_*=1.8$.}
\label{tab:f12g18-ffV}
\end{table}
%%%%%%%%%%%%%%%%%%%%%%%%%%%%%%%%%%%%%%%%%%%%%v
%%% ACCOPPIAMENTI DI GAUGE
%%%%%%%%%%%%%%%%%%%%%%%%%%%%%%%%%%%%%%%%%%%%%%
\begin{table}[!h]
%\captionsetup[subfloat]{labelformat=empty,position=top}
\centering
\begin{tabular}{||c||c|c|c|c||}
\hline\hline
$Z$         & $W^{+}$         &$W_2^{+}$         &$W_3^{+}$        \\
\hline
\hline
$W^{-}$        & $-0.57$         &        0.002         & $-2.3\cdot10^{-6}$        \\
%\hline
$W_2^{-}$        &        0.002         &        $-0.56$         &$-0.034$        \\
%\hline
$W_3^{-}$        &$-2.3\cdot10^{-6}$        &        $-0.034$         &$-0.20$        \\
\hline\hline
\end{tabular}
\begin{tabular}{||c||c|c|c|c||}
\hline\hline
$Z_2$         & $W^{+}$         &$W_2^{+}$         &$W_3^{+}$        \\
\hline
\hline
$W^{-}$        &$-0.001$        &        $-0.018$ &        0.018        \\
%\hline
$W_2^{-}$        &$-0.018$        &        $-0.042$ &        $-0.006$        \\
%\hline
$W_3^{-}$        &0.018        &        $-0.006$ &        0.58        \\
\hline\hline
\end{tabular}\\
\vspace{0.1cm}
\begin{tabular}{||c||c|c|c|c||}
\hline\hline
$Z_3$         & $W^{+}$         &$W_2^{+}$         &$W_3^{+}$        \\
\hline
\hline
$W^{-}$        &0.001        &         $-0.64$        &        $-0.030$        \\
%\hline
$W_2^{-}$        &$-0.64$        &        $-1.4$ &        0.023\\
%\hline
$W_3^{-}$        &$-0.030$        &        0.023 &        $-0.86$        \\
\hline\hline
\end{tabular}
\begin{tabular}{||c||c|c|c|c||}
\hline\hline
$Z_5$         & $W^{+}$         &$W_2^{+}$         &$W_3^{+}$        \\
\hline
\hline
$W^{-}$        &$-5.9\cdot10^{-6}$        &         $-0.030$        &        $-0.33$         \\
%\hline
$W_2^{-}$        &$-0.030$         &         0.023        &        $-0.89$        \\
%\hline
$W_3^{-}$        &$-0.33$         &         $-0.89$        &$-0.031$        \\
\hline\hline
\end{tabular}
\caption{Tri-linear coupling of the neutral to the charged gauge bosons of the 4DCHM for the case $f=1.2 $ TeV and $g_*=1.8$.}
\label{tab:f12g18-VVV}
\end{table}
\clearpage
\bibliography{bibliography}

\hyphenation{Post-Script Sprin-ger}
\begin{thebibliography}{42}
\expandafter\ifx\csname natexlab\endcsname\relax\def\natexlab#1{#1}\fi
\expandafter\ifx\csname bibnamefont\endcsname\relax
  \def\bibnamefont#1{#1}\fi
\expandafter\ifx\csname bibfnamefont\endcsname\relax
  \def\bibfnamefont#1{#1}\fi
\expandafter\ifx\csname citenamefont\endcsname\relax
  \def\citenamefont#1{#1}\fi
\expandafter\ifx\csname url\endcsname\relax
  \def\url#1{\texttt{#1}}\fi
\expandafter\ifx\csname urlprefix\endcsname\relax\def\urlprefix{URL }\fi
\providecommand{\bibinfo}[2]{#2}
\providecommand{\eprint}[2][]{\url{#2}}

\bibitem[{\citenamefont{Aad et~al.}(2012{\natexlab{a}})}]{Aad:2012rga}
\bibinfo{author}{\bibfnamefont{G.}~\bibnamefont{Aad}} \bibnamefont{et~al.}
  (\bibinfo{collaboration}{ATLAS Collaboration})
  (\bibinfo{year}{2012}{\natexlab{a}}), \eprint{1208.1390}.

\bibitem[{\citenamefont{Chatrchyan
  et~al.}(2012{\natexlab{a}})}]{Chatrchyan:2012bd}
\bibinfo{author}{\bibfnamefont{S.}~\bibnamefont{Chatrchyan}}
  \bibnamefont{et~al.} (\bibinfo{collaboration}{CMS Collaboration})
  (\bibinfo{year}{2012}{\natexlab{a}}), \eprint{1210.7544}.

\bibitem[{\citenamefont{Campbell et~al.}(2007)\citenamefont{Campbell, Huston,
  and Stirling}}]{Campbell:2006wx}
\bibinfo{author}{\bibfnamefont{J.~M.} \bibnamefont{Campbell}},
  \bibinfo{author}{\bibfnamefont{J.}~\bibnamefont{Huston}}, \bibnamefont{and}
  \bibinfo{author}{\bibfnamefont{W.}~\bibnamefont{Stirling}},
  \bibinfo{journal}{Rept. Prog. Phys.} \textbf{\bibinfo{volume}{70}},
  \bibinfo{pages}{89} (\bibinfo{year}{2007}), \eprint{hep-ph/0611148}.

\bibitem[{\citenamefont{Aad et~al.}(2012{\natexlab{b}})}]{:2012gk}
\bibinfo{author}{\bibfnamefont{G.}~\bibnamefont{Aad}} \bibnamefont{et~al.}
  (\bibinfo{collaboration}{ATLAS Collaboration}), \bibinfo{journal}{Phys.
  Lett.} \textbf{\bibinfo{volume}{B716}}, \bibinfo{pages}{1}
  (\bibinfo{year}{2012}{\natexlab{b}}), \eprint{1207.7214}.

\bibitem[{\citenamefont{Chatrchyan et~al.}(2012{\natexlab{b}})}]{:2012gu}
\bibinfo{author}{\bibfnamefont{S.}~\bibnamefont{Chatrchyan}}
  \bibnamefont{et~al.} (\bibinfo{collaboration}{CMS Collaboration}),
  \bibinfo{journal}{Phys. Lett.} \textbf{\bibinfo{volume}{B716}},
  \bibinfo{pages}{30} (\bibinfo{year}{2012}{\natexlab{b}}), \eprint{1207.7235}.

\bibitem[{\citenamefont{De~Curtis
  et~al.}(2012{\natexlab{a}})\citenamefont{De~Curtis, Redi, and
  Tesi}}]{DeCurtis:2011yx}
\bibinfo{author}{\bibfnamefont{S.}~\bibnamefont{De~Curtis}},
  \bibinfo{author}{\bibfnamefont{M.}~\bibnamefont{Redi}}, \bibnamefont{and}
  \bibinfo{author}{\bibfnamefont{A.}~\bibnamefont{Tesi}},
  \bibinfo{journal}{JHEP} \textbf{\bibinfo{volume}{1204}}, \bibinfo{pages}{042}
  (\bibinfo{year}{2012}{\natexlab{a}}), \eprint{1110.1613}.

\bibitem[{\citenamefont{Kaplan and Georgi}(1984)}]{Kaplan:1983fs}
\bibinfo{author}{\bibfnamefont{D.~B.} \bibnamefont{Kaplan}} \bibnamefont{and}
  \bibinfo{author}{\bibfnamefont{H.}~\bibnamefont{Georgi}},
  \bibinfo{journal}{Phys. Lett.} \textbf{\bibinfo{volume}{B136}},
  \bibinfo{pages}{183} (\bibinfo{year}{1984}).

\bibitem[{\citenamefont{Georgi et~al.}(1984)\citenamefont{Georgi, Kaplan, and
  Galison}}]{Georgi:1984ef}
\bibinfo{author}{\bibfnamefont{H.}~\bibnamefont{Georgi}},
  \bibinfo{author}{\bibfnamefont{D.~B.} \bibnamefont{Kaplan}},
  \bibnamefont{and} \bibinfo{author}{\bibfnamefont{P.}~\bibnamefont{Galison}},
  \bibinfo{journal}{Phys. Lett.} \textbf{\bibinfo{volume}{B143}},
  \bibinfo{pages}{152} (\bibinfo{year}{1984}).

\bibitem[{\citenamefont{Georgi and Kaplan}(1984)}]{Georgi:1984af}
\bibinfo{author}{\bibfnamefont{H.}~\bibnamefont{Georgi}} \bibnamefont{and}
  \bibinfo{author}{\bibfnamefont{D.~B.} \bibnamefont{Kaplan}},
  \bibinfo{journal}{Phys. Lett.} \textbf{\bibinfo{volume}{B145}},
  \bibinfo{pages}{216} (\bibinfo{year}{1984}).

\bibitem[{\citenamefont{Dugan et~al.}(1985)\citenamefont{Dugan, Georgi, and
  Kaplan}}]{Dugan:1984hq}
\bibinfo{author}{\bibfnamefont{M.~J.} \bibnamefont{Dugan}},
  \bibinfo{author}{\bibfnamefont{H.}~\bibnamefont{Georgi}}, \bibnamefont{and}
  \bibinfo{author}{\bibfnamefont{D.~B.} \bibnamefont{Kaplan}},
  \bibinfo{journal}{Nucl. Phys.} \textbf{\bibinfo{volume}{B254}},
  \bibinfo{pages}{299} (\bibinfo{year}{1985}).

\bibitem[{\citenamefont{Agashe et~al.}(2005)\citenamefont{Agashe, Contino, and
  Pomarol}}]{Agashe:2004rs}
\bibinfo{author}{\bibfnamefont{K.}~\bibnamefont{Agashe}},
  \bibinfo{author}{\bibfnamefont{R.}~\bibnamefont{Contino}}, \bibnamefont{and}
  \bibinfo{author}{\bibfnamefont{A.}~\bibnamefont{Pomarol}},
  \bibinfo{journal}{Nucl. Phys.} \textbf{\bibinfo{volume}{B719}},
  \bibinfo{pages}{165} (\bibinfo{year}{2005}), \eprint{hep-ph/0412089}.

\bibitem[{\citenamefont{Matsedonskyi et~al.}(2012)\citenamefont{Matsedonskyi,
  Panico, and Wulzer}}]{Matsedonskyi:2012ym}
\bibinfo{author}{\bibfnamefont{O.}~\bibnamefont{Matsedonskyi}},
  \bibinfo{author}{\bibfnamefont{G.}~\bibnamefont{Panico}}, \bibnamefont{and}
  \bibinfo{author}{\bibfnamefont{A.}~\bibnamefont{Wulzer}}
  (\bibinfo{year}{2012}), \eprint{1204.6333}.

\bibitem[{\citenamefont{Pomarol and Riva}(2012)}]{Pomarol:2012qf}
\bibinfo{author}{\bibfnamefont{A.}~\bibnamefont{Pomarol}} \bibnamefont{and}
  \bibinfo{author}{\bibfnamefont{F.}~\bibnamefont{Riva}},
  \bibinfo{journal}{JHEP} \textbf{\bibinfo{volume}{1208}}, \bibinfo{pages}{135}
  (\bibinfo{year}{2012}), \eprint{1205.6434}.

\bibitem[{\citenamefont{Redi and Tesi}(2012)}]{Redi:2012ha}
\bibinfo{author}{\bibfnamefont{M.}~\bibnamefont{Redi}} \bibnamefont{and}
  \bibinfo{author}{\bibfnamefont{A.}~\bibnamefont{Tesi}}
  (\bibinfo{year}{2012}), \eprint{1205.0232}.

\bibitem[{\citenamefont{Marzocca et~al.}(2012)\citenamefont{Marzocca, Serone,
  and Shu}}]{Marzocca:2012zn}
\bibinfo{author}{\bibfnamefont{D.}~\bibnamefont{Marzocca}},
  \bibinfo{author}{\bibfnamefont{M.}~\bibnamefont{Serone}}, \bibnamefont{and}
  \bibinfo{author}{\bibfnamefont{J.}~\bibnamefont{Shu}},
  \bibinfo{journal}{JHEP} \textbf{\bibinfo{volume}{1208}}, \bibinfo{pages}{013}
  (\bibinfo{year}{2012}), \eprint{1205.0770}.

\bibitem[{\citenamefont{Panico et~al.}(2012)\citenamefont{Panico, Redi, Tesi,
  and Wulzer}}]{Panico:2012uw}
\bibinfo{author}{\bibfnamefont{G.}~\bibnamefont{Panico}},
  \bibinfo{author}{\bibfnamefont{M.}~\bibnamefont{Redi}},
  \bibinfo{author}{\bibfnamefont{A.}~\bibnamefont{Tesi}}, \bibnamefont{and}
  \bibinfo{author}{\bibfnamefont{A.}~\bibnamefont{Wulzer}}
  (\bibinfo{year}{2012}), \eprint{1210.7114}.

\bibitem[{\citenamefont{Redi}(2011)}]{RediTalk}
\bibinfo{author}{\bibfnamefont{M.}~\bibnamefont{Redi}} (\bibinfo{year}{2011}),
  \eprint{talk given at the `Interpreting LHC Discoveries Conference', GGI,
  8-11 November 2011, see http://www.ggi.fi.infn.it/talks/talk2175.pdf}.

\bibitem[{\citenamefont{Barducci
  et~al.}(2012{\natexlab{a}})\citenamefont{Barducci, De~Curtis, Mimasu,
  Moretti, and Pruna}}]{Ken:2012pp}
\bibinfo{author}{\bibfnamefont{D.}~\bibnamefont{Barducci}},
  \bibinfo{author}{\bibfnamefont{S.}~\bibnamefont{De~Curtis}},
  \bibinfo{author}{\bibfnamefont{K.}~\bibnamefont{Mimasu}},
  \bibinfo{author}{\bibfnamefont{S.}~\bibnamefont{Moretti}}, \bibnamefont{and}
  \bibinfo{author}{\bibfnamefont{G.~M.} \bibnamefont{Pruna}}
  (\bibinfo{year}{2012}{\natexlab{a}}), \eprint{in progress}.

\bibitem[{\citenamefont{Barducci
  et~al.}(2012{\natexlab{b}})\citenamefont{Barducci, Belyaev, De~Curtis,
  Moretti, and Pruna}}]{Barducci:2012kk}
\bibinfo{author}{\bibfnamefont{D.}~\bibnamefont{Barducci}},
  \bibinfo{author}{\bibfnamefont{A.}~\bibnamefont{Belyaev}},
  \bibinfo{author}{\bibfnamefont{S.}~\bibnamefont{De~Curtis}},
  \bibinfo{author}{\bibfnamefont{S.}~\bibnamefont{Moretti}}, \bibnamefont{and}
  \bibinfo{author}{\bibfnamefont{G.~M.} \bibnamefont{Pruna}}
  (\bibinfo{year}{2012}{\natexlab{b}}), \eprint{1210.2927}.

\bibitem[{\citenamefont{De~Curtis
  et~al.}(2012{\natexlab{b}})\citenamefont{De~Curtis, Dominici, Fedeli, and
  Moretti}}]{DeCurtis:2012cn}
\bibinfo{author}{\bibfnamefont{S.}~\bibnamefont{De~Curtis}},
  \bibinfo{author}{\bibfnamefont{D.}~\bibnamefont{Dominici}},
  \bibinfo{author}{\bibfnamefont{L.}~\bibnamefont{Fedeli}}, \bibnamefont{and}
  \bibinfo{author}{\bibfnamefont{S.}~\bibnamefont{Moretti}}
  (\bibinfo{year}{2012}{\natexlab{b}}), \eprint{1210.7649}.

\bibitem[{\citenamefont{Aad et~al.}(2011)}]{Aad:2011fe}
\bibinfo{author}{\bibfnamefont{G.}~\bibnamefont{Aad}} \bibnamefont{et~al.}
  (\bibinfo{collaboration}{ATLAS Collaboration}), \bibinfo{journal}{Phys.
  Lett.} \textbf{\bibinfo{volume}{B701}}, \bibinfo{pages}{50}
  (\bibinfo{year}{2011}), \eprint{1103.1391}.

\bibitem[{\citenamefont{Chatrchyan
  et~al.}(2012{\natexlab{c}})}]{Chatrchyan:2012qk}
\bibinfo{author}{\bibfnamefont{S.}~\bibnamefont{Chatrchyan}}
  \bibnamefont{et~al.} (\bibinfo{collaboration}{CMS Collaboration})
  (\bibinfo{year}{2012}{\natexlab{c}}), \eprint{1204.4764}.

\bibitem[{\citenamefont{Hayden et~al.}(2012)}]{Hayden:2012gc}
\bibinfo{author}{\bibfnamefont{D.}~\bibnamefont{Hayden}} \bibnamefont{et~al.}
  (\bibinfo{collaboration}{ATLAS Collaboration}), \bibinfo{journal}{EPJ Web
  Conf.} \textbf{\bibinfo{volume}{28}}, \bibinfo{pages}{12003}
  (\bibinfo{year}{2012}), \eprint{1201.4721}.

\bibitem[{\citenamefont{Chatrchyan
  et~al.}(2012{\natexlab{d}})}]{Chatrchyan:2012it}
\bibinfo{author}{\bibfnamefont{S.}~\bibnamefont{Chatrchyan}}
  \bibnamefont{et~al.} (\bibinfo{collaboration}{CMS Collaboration})
  (\bibinfo{year}{2012}{\natexlab{d}}), \eprint{1206.1849}.

\bibitem[{\citenamefont{Aad et~al.}(2012{\natexlab{c}})}]{Aad:2012bb}
\bibinfo{author}{\bibfnamefont{G.}~\bibnamefont{Aad}} \bibnamefont{et~al.}
  (\bibinfo{collaboration}{ATLAS Collaboration}), \bibinfo{journal}{JHEP}
  \textbf{\bibinfo{volume}{1204}}, \bibinfo{pages}{069}
  (\bibinfo{year}{2012}{\natexlab{c}}), \eprint{1202.5520}.

\bibitem[{\citenamefont{Aad et~al.}(2012{\natexlab{d}})}]{ATLAS:2012aw}
\bibinfo{author}{\bibfnamefont{G.}~\bibnamefont{Aad}} \bibnamefont{et~al.}
  (\bibinfo{collaboration}{ATLAS Collaboration}), \bibinfo{journal}{Phys. Rev.
  Lett.} \textbf{\bibinfo{volume}{109}}, \bibinfo{pages}{032001}
  (\bibinfo{year}{2012}{\natexlab{d}}), \eprint{1202.6540}.

\bibitem[{\citenamefont{Chatrchyan
  et~al.}(2012{\natexlab{e}})}]{Chatrchyan:2012fp}
\bibinfo{author}{\bibfnamefont{S.}~\bibnamefont{Chatrchyan}}
  \bibnamefont{et~al.} (\bibinfo{collaboration}{CMS Collaboration})
  (\bibinfo{year}{2012}{\natexlab{e}}), \eprint{1209.1062}.

\bibitem[{\citenamefont{Contino et~al.}(2007)\citenamefont{Contino, Da~Rold,
  and Pomarol}}]{Contino:2006qr}
\bibinfo{author}{\bibfnamefont{R.}~\bibnamefont{Contino}},
  \bibinfo{author}{\bibfnamefont{L.}~\bibnamefont{Da~Rold}}, \bibnamefont{and}
  \bibinfo{author}{\bibfnamefont{A.}~\bibnamefont{Pomarol}},
  \bibinfo{journal}{Phys. Rev.} \textbf{\bibinfo{volume}{D75}},
  \bibinfo{pages}{055014} (\bibinfo{year}{2007}), \eprint{hep-ph/0612048}.

\bibitem[{\citenamefont{Accomando et~al.}(2012)\citenamefont{Accomando, Fedeli,
  Moretti, De~Curtis, and Dominici}}]{Accomando:2012yg}
\bibinfo{author}{\bibfnamefont{E.}~\bibnamefont{Accomando}},
  \bibinfo{author}{\bibfnamefont{L.}~\bibnamefont{Fedeli}},
  \bibinfo{author}{\bibfnamefont{S.}~\bibnamefont{Moretti}},
  \bibinfo{author}{\bibfnamefont{S.}~\bibnamefont{De~Curtis}},
  \bibnamefont{and} \bibinfo{author}{\bibfnamefont{D.}~\bibnamefont{Dominici}}
  (\bibinfo{year}{2012}), \eprint{1208.0268}.

\bibitem[{\citenamefont{Semenov}(1996)}]{Semenov:1996es}
\bibinfo{author}{\bibfnamefont{A.~V.} \bibnamefont{Semenov}}
  (\bibinfo{year}{1996}), \eprint{hep-ph/9608488}.

\bibitem[{\citenamefont{Semenov}(2010)}]{Semenov:2010qt}
\bibinfo{author}{\bibfnamefont{A.}~\bibnamefont{Semenov}}
  (\bibinfo{year}{2010}), \eprint{1005.1909}.

\bibitem[{\citenamefont{Pukhov}(2004)}]{Pukhov:2004ca}
\bibinfo{author}{\bibfnamefont{A.}~\bibnamefont{Pukhov}}
  (\bibinfo{year}{2004}), \eprint{hep-ph/0412191}.

\bibitem[{\citenamefont{Belyaev et~al.}(2012)\citenamefont{Belyaev,
  Christensen, and Pukhov}}]{Belyaev:2012qa}
\bibinfo{author}{\bibfnamefont{A.}~\bibnamefont{Belyaev}},
  \bibinfo{author}{\bibfnamefont{N.~D.} \bibnamefont{Christensen}},
  \bibnamefont{and} \bibinfo{author}{\bibfnamefont{A.}~\bibnamefont{Pukhov}}
  (\bibinfo{year}{2012}), \eprint{1207.6082}.

\bibitem[{\citenamefont{Bondarenko et~al.}(2012)}]{Brooijmans:2012yi}
\bibinfo{author}{\bibfnamefont{M.}~\bibnamefont{Bondarenko}}
  \bibnamefont{et~al.} (\bibinfo{year}{2012}), \eprint{1203.1488}.

\bibitem[{\citenamefont{Ballestrero}(1999)}]{Ballestrero:1999md}
\bibinfo{author}{\bibfnamefont{A.}~\bibnamefont{Ballestrero}}, pp.
  \bibinfo{pages}{303--309} (\bibinfo{year}{1999}), \eprint{hep-ph/9911318}.

\bibitem[{\citenamefont{Murayama et~al.}(1992)\citenamefont{Murayama, Watanabe,
  and Hagiwara}}]{Murayama:1992gi}
\bibinfo{author}{\bibfnamefont{H.}~\bibnamefont{Murayama}},
  \bibinfo{author}{\bibfnamefont{I.}~\bibnamefont{Watanabe}}, \bibnamefont{and}
  \bibinfo{author}{\bibfnamefont{K.}~\bibnamefont{Hagiwara}},
  \emph{\bibinfo{title}{{HELAS: HELicity amplitude subroutines for Feynman
  diagram evaluations}}} (\bibinfo{year}{1992}).

\bibitem[{\citenamefont{Stelzer and Long}(1994)}]{Stelzer:1994ta}
\bibinfo{author}{\bibfnamefont{T.}~\bibnamefont{Stelzer}} \bibnamefont{and}
  \bibinfo{author}{\bibfnamefont{W.}~\bibnamefont{Long}},
  \bibinfo{journal}{Comput. Phys. Commun.} \textbf{\bibinfo{volume}{81}},
  \bibinfo{pages}{357} (\bibinfo{year}{1994}), \eprint{hep-ph/9401258}.

\bibitem[{\citenamefont{Kleiss et~al.}(1986)\citenamefont{Kleiss, Stirling, and
  Ellis}}]{Kleiss:1985gy}
\bibinfo{author}{\bibfnamefont{R.}~\bibnamefont{Kleiss}},
  \bibinfo{author}{\bibfnamefont{W.~J.} \bibnamefont{Stirling}},
  \bibnamefont{and} \bibinfo{author}{\bibfnamefont{S.}~\bibnamefont{Ellis}},
  \bibinfo{journal}{Comput. Phys. Commun.} \textbf{\bibinfo{volume}{40}},
  \bibinfo{pages}{359} (\bibinfo{year}{1986}).

\bibitem[{\citenamefont{Lepage}(1978)}]{Lepage:1977sw}
\bibinfo{author}{\bibfnamefont{G.~P.} \bibnamefont{Lepage}},
  \bibinfo{journal}{J. Comput. Phys.} \textbf{\bibinfo{volume}{27}},
  \bibinfo{pages}{192} (\bibinfo{year}{1978}).

\bibitem[{\citenamefont{Lai et~al.}(2000)}]{Lai:1999wy}
\bibinfo{author}{\bibfnamefont{H.}~\bibnamefont{Lai}} \bibnamefont{et~al.}
  (\bibinfo{collaboration}{CTEQ Collaboration}), \bibinfo{journal}{Eur. Phys.
  J.} \textbf{\bibinfo{volume}{C12}}, \bibinfo{pages}{375}
  (\bibinfo{year}{2000}), \eprint{hep-ph/9903282}.

\bibitem[{\citenamefont{Aad et~al.}(2012{\natexlab{e}})}]{AA:2012ks}
\bibinfo{author}{\bibfnamefont{G.}~\bibnamefont{Aad}} \bibnamefont{et~al.}
  (\bibinfo{collaboration}{ATLAS Collaboration}), \bibinfo{journal}{Phys.
  Lett.} \textbf{\bibinfo{volume}{B712}}, \bibinfo{pages}{289}
  (\bibinfo{year}{2012}{\natexlab{e}}), \eprint{1203.6232}.

\bibitem[{\citenamefont{Gianotti et~al.}(2005)\citenamefont{Gianotti, Mangano,
  Virdee, Abdullin, Azuelos et~al.}}]{Gianotti:2002xx}
\bibinfo{author}{\bibfnamefont{F.}~\bibnamefont{Gianotti}},
  \bibinfo{author}{\bibfnamefont{M.}~\bibnamefont{Mangano}},
  \bibinfo{author}{\bibfnamefont{T.}~\bibnamefont{Virdee}},
  \bibinfo{author}{\bibfnamefont{S.}~\bibnamefont{Abdullin}},
  \bibinfo{author}{\bibfnamefont{G.}~\bibnamefont{Azuelos}},
  \bibnamefont{et~al.}, \bibinfo{journal}{Eur. Phys. J.}
  \textbf{\bibinfo{volume}{C39}}, \bibinfo{pages}{293} (\bibinfo{year}{2005}),
  \eprint{hep-ph/0204087}.

\end{thebibliography}
\end{document}